\pdfoutput=1

\documentclass[11pt,twoside,a4paper,cmspaper,final,collab]{cms-tdr}
\def\svnVersion{370072}\def\cmsCernNoTag{CERN-EP/2016-050}\def\cmsCernDate{\today}\def\cmsMessage{Published in Physical Review D as \href{http://dx.doi.org/10.1103/PhysRevD.94.052012}{\doi{10.1103/PhysRevD.94.052012.}}}

\begin{document}\cmsNoteHeader{HIG-13-032}

\hyphenation{had-ron-i-za-tion}
\hyphenation{cal-or-i-me-ter}
\hyphenation{de-vices}
\RCS$HeadURL: svn+ssh://svn.cern.ch/reps/tdr2/papers/HIG-13-032/trunk/HIG-13-032.tex $
\RCS$Id: HIG-13-032.tex 370015 2016-10-06 09:52:28Z mgouzevi $
\newlength\cmsFigWidth
\ifthenelse{\boolean{cms@external}}{\setlength\cmsFigWidth{0.48\textwidth}}{\setlength\cmsFigWidth{0.6\textwidth}}
\ifthenelse{\boolean{cms@external}}{\providecommand{\cmsLeft}{top\xspace}}{\providecommand{\cmsLeft}{left\xspace}}
\ifthenelse{\boolean{cms@external}}{\providecommand{\cmsRight}{bottom\xspace}}{\providecommand{\cmsRight}{right\xspace}}
\ifthenelse{\boolean{cms@external}}{\providecommand{\CL}{C.L.\xspace}}{\providecommand{\CL}{CL\xspace}}
\newlength\cmsTabSkip\setlength\cmsTabSkip{1.5ex}

\newcommand{\sggG}{\ensuremath{\sigma_{\gamma \gamma}^\text{G}}\xspace}
\newcommand{\sjjG}{\ensuremath{\sigma_\mathrm{jj}^\text{G}}\xspace}
\newcommand{\sggjjkG}{\ensuremath{\sigma_{\gamma \gamma \mathrm{jj}}^\text{G, kin}}\xspace}
\newcommand{\sggCB}{\ensuremath{\sigma_{\gamma \gamma}^\text{CB}}\xspace}
\newcommand{\sjjCB}{\ensuremath{\sigma_\mathrm{jj}^\text{CB}}\xspace}
\newcommand{\sggjjkCB}{\ensuremath{\sigma_{\gamma \gamma \mathrm{jj}}^\text{CB, kin}}\xspace}
\newcommand{\pTg}{\ensuremath{p_\mathrm{T}^{\gamma}}\xspace}
\newcommand{\pTgone}{\ensuremath{p_\mathrm{T}^{\gamma 1}}\xspace}
\newcommand{\pTgtwo}{\ensuremath{p_\mathrm{T}^{\gamma 2}}\xspace}
\newcommand{\HH}{\ensuremath{\PH\PH}\xspace}
\newcommand{\Mgg}{\ensuremath{m_{\gamma\gamma}}\xspace}
\newcommand{\Mjj}{\ensuremath{m_\mathrm{jj}}\xspace}
\newcommand{\Mugg}{\ensuremath{\mu_{\gamma\gamma}}\xspace}
\newcommand{\Mujj}{\ensuremath{\mu_\mathrm{jj}}\xspace}
\newcommand{\thetastar}{\ensuremath{\theta^\mathrm{CS}_{\rm \PH\PH}}\xspace}
\newcommand{\acosthetastar}{\ensuremath{\bigl|\mathrm{cos}\,\theta^\mathrm{CS}_{\PH\PH}\bigr|}\xspace}
\newcommand{\Mggjj}{\ensuremath{m_{\gamma\gamma \mathrm{jj}}}\xspace}
\newcommand{\Mggjjk}{\ensuremath{m_{\gamma\gamma \mathrm{jj}}^\text{kin}}\xspace}
\newcommand{\Muggjjk}{\ensuremath{\mu_{\gamma\gamma \mathrm{jj}}^\text{kin}}\xspace}
\newcommand{\pT}{\pt}
\newcommand{\pTj}{\ensuremath{p_\mathrm{T}^\mathrm{j}}\xspace}
\newcommand{\etaj}{\ensuremath{\eta_\mathrm{j}}\xspace}
\newcommand{\kapt}{\ensuremath{\kappa_{\PQt}}\xspace}
\newcommand{\kapl}{\ensuremath{\kappa_{\lambda}}\xspace}
\newcommand{\mH}{\ensuremath{m_{\PH}}\xspace}
\newcommand{\sigmaHH}{\ensuremath{\sigma_{\PH\PH}}\xspace}
\newcommand{\mx}{\ensuremath{m_\mathrm{X}}\xspace}
\newcommand{\ctwo}{\ensuremath{c_2}\xspace}
\newcommand{\yt}{\ensuremath{y_{\PQt}}\xspace}
\newcommand{\mt}{\ensuremath{m_{\PQt}}\xspace}
\newcommand{\lbdSM}{\ensuremath{\lambda^\mathrm{SM}}\xspace}
\newcommand{\DRgj}{\ensuremath{\Delta R_{\gamma \mathrm{j}}}\xspace}
\newcommand{\LambdaR}{\ensuremath{\Lambda_\mathrm{R}}\xspace}
\newcommand{\AMpl}{\ensuremath{\overline{M}_\mathrm{Pl}}\xspace}

\newcommand\tab[1][1cm]{\hspace*{#1}}

\cmsNoteHeader{HIG-13-032}
\title{Search for two Higgs bosons in final states containing two photons and two bottom quarks in proton-proton collisions at 8 TeV}

\date{\today}

\abstract{
A search is presented for the production of two Higgs bosons in final states containing two photons and two bottom quarks.
Both resonant and nonresonant hypotheses are investigated.
The analyzed data correspond to an integrated luminosity of 19.7\fbinv of proton-proton collisions at $\sqrt{s}=8$\TeV collected
with the CMS detector.
Good agreement is observed between data and predictions of the
standard model (SM).
Upper limits are set at 95\% confidence level on the production
cross section of new particles and compared to the prediction
for the existence of a warped extra dimension.
When the decay to two Higgs bosons is kinematically allowed, assuming a mass scale $\LambdaR = 1$\TeV for the model, the data exclude a radion scalar
at masses below 980\GeV.
The first Kaluza--Klein excitation mode of the graviton in the RS1 Randall--Sundrum model is excluded for masses between 325 and 450\GeV.
An upper limit of 0.71\unit{pb} is set on the nonresonant two-Higgs boson cross section in the SM-like hypothesis. Limits are also derived on nonresonant
production assuming anomalous Higgs boson couplings.
}

\hypersetup{%
pdfauthor={CMS Collaboration},%
pdftitle={Search for two Higgs bosons in final states containing two photons and two bottom quarks},%
pdfsubject={CMS},%
pdfkeywords={CMS, BSM physics, Higgs}}

\maketitle

\section{Introduction}

The discovery of a boson with a mass of approximately 125\GeV, with properties close to those expected for the
Higgs boson (\PH) of the standard model (SM)~\cite{Chatrchyan:2012ufa,HiggsdiscoveryAtlas}, has stimulated
interest in the exploration of the Higgs potential.
The production of a pair of Higgs bosons ($\PH\PH$)  is a rare process that is sensitive to the structure of this potential through the self-coupling mechanism of the Higgs boson.
In the SM, the cross section for the production of two Higgs bosons in $\Pp\Pp$ collisions at 8\TeV is $10.0 \pm 1.4\unit{fb}$
for the gluon-gluon fusion process~\cite{deFlorian:2013jea, Dawson:1998py, Baglio:2012np}, which lies beyond the reach of analyses based on the first run of
the CERN LHC.

Many theories beyond the SM (BSM) suggest the existence of heavy particles that can couple to a pair of Higgs bosons. These particles could appear as a resonant contribution in the invariant mass of the HH system. If the new particles are too heavy to be observed through a direct search, they may be sensed in the HH production through their virtual contributions (as shown, e.g., in Refs.~\cite{Dawson:2015oha,Heng:2013cya}); also, the fundamental couplings of the model can be modified relative to their SM values (as shown, e.g., in Refs.~\cite{Grober:2010yv,Moretti:2004wa}); in both cases, a nonresonant enhancement of the HH production could be observed.

Models with a warped extra dimension
(WED), as proposed by Randall and Sundrum~\cite{Randall:1999ee}, postulate the existence of one spatial extra dimension compactified between two fixed points, commonly called branes. The region between the branes is referred to as bulk, and controlled through an exponential metric.
The gap between the two fundamental scales of nature, such as the Planck scale (\Mpl), and the electroweak scale, is controlled by a warp factor ($k$) in the metric, which corresponds to one of the fundamental parameters of  the model. The brane where the density of the extra dimensional metric is localized is called ``Planck brane", while the other, where the Higgs field is localized, is called ``\TeVns{} brane".
This class of models predicts the existence of new particles that can decay to Higgs boson pair, such as the spin-0 radion~\cite{Goldberger:1999uk,DeWolfe:1999cp,Csaki:1999mp}, and the spin-2
first Kaluza--Klein (KK) excitation of the graviton~\cite{Davoudiasl:1999jd,Csaki:2000zn, Agashe:2007zd}.

There are two possible ways of describing a KK graviton in WED that depend on the choice of localization for the SM matter fields.
In the RS1 model, only gravity is allowed to propagate in the extra-dimensional bulk. In this model the couplings of the KK-graviton to matter fields are controlled by $k/\AMpl$~\cite{Randall:1999ee}, with the reduced Planck mass $\AMpl$ defined by $\Mpl /\sqrt{8\pi}$. For the possibility of SM particles to propagate in the bulk (the so-called bulk-RS model), the coupling of the KK graviton to matter depends on the choice for the localization of the SM bulk fields. This paper uses the phenomenology of Ref. \cite{Fitzpatrick:2007qr}, where SM particles are allowed to propagate in the bulk, and follows the characteristics of the SM gauge group, with the right-handed top quark localized on the TeV brane (so called elementary top hypothesis).

The radion (R) is an additional element of WED models that is needed to stabilize the size of the extra dimension $l$. It is usual to express the benchmark points of the model in terms of the dimensionless quantity $k/\AMpl$, and the mass scale $\LambdaR = \sqrt{6} \exp[-kl]\AMpl$, with the latter interpreted as the ultraviolet cutoff of the model~\cite{Giudice:2000av}.
The addition of a scalar-curvature term can induce a mixing between the
scalar radion and the Higgs boson~\cite{Giudice:2000av,Dominici:2002jv}. This possibility is discussed, for example, in Ref.~\cite{Antoniadis:2002ut}.
Precision electroweak studies suggest that this mixing is expected to be small~\cite{Desai:2013pga}.
In our interpretations of the constraints we neglect the possibility of Higgs--radion mixing.

On one hand, the choice of localization of the SM matter fields for the KK-graviton resonance impacts the kinematics
of the signal and drastically modifies the production and decay properties~\cite{Oliveira:2010uv}.
The physics of the radion, on the other hand, does not depend much on the choice of the model~\cite{Giudice:2000av}, which obviates the need to distinguish the RS1 and bulk-RS possibilities.

Models with an extended Higgs sector also predict one spin-0 resonance that, when sufficiently massive, decays to a pair of SM Higgs bosons, and would correspond to an additional Higgs boson.
Examples of such models are the singlet extension~\cite{O'Connell:2006wi}, the two Higgs doublet models~\cite{Branco:2011iw} (in particular the minimal supersymmetric model ~\cite{Djouadi:2005gj,Barbieri:2013nka}), and the Georgi-Machacek model~\cite{Georgi:1985nv}. The majority of these models predict that heavy scalar production occurs predominantly through the gluon-gluon fusion process. The Lorentz structure of the coupling between the scalar and the gluon is the same for a radion or a heavy Higgs boson. Therefore the models for the production of a radion or an additional Higgs boson are essentially the same, provided the interpretations are performed in a parameter space region where the spin-0 resonance is narrow. The results of this paper can therefore be easily applied to constrain this class of models.

Phenomenological explorations  of the two-Higgs-boson channel were studied prior to the observation of the Higgs boson~\cite{Glover:1987nx}, and, since then, other studies have become available~\cite{Plehn:1996wb,Djouadi:1999rca,Baur:2003gp,Pierce:2006dh,Nishiwaki:2013cma,Goertz:2014qta,Liu:2014rba}. Most of these indicate that in BSM physics an enhancement of the $\HH$ production cross section is expected, together with modified signal kinematics for the $\HH$ final state.
This paper describes a search for the production of pairs of Higgs bosons in the $\gamma\gamma \bbbar$
final state in proton-proton (pp) collisions at the LHC,
using data corresponding to an integrated luminosity of 19.7\fbinv collected by the CMS experiment at $\sqrt{s}=8\TeV$.
Both nonresonant and resonant production are explored, with the search for a narrow resonance X conducted at masses $\mx$ between 260 and 1100\GeV.

The fully-reconstructed $\gamma\gamma \bbbar$ final state discussed in this paper,
combines the large SM branching fraction ($\mathcal{B}$) of the $\PH \to \bbbar$ decay
with the comparatively low background and good mass resolution of the $\PH \to \gamma \gamma$ channel,
yielding a total $\mathcal{B}(\HH \to \gamma\gamma \bbbar)$ of 0.26\%~\cite{Heinemeyer:2013tqa}.
The search exploits the mass spectra of the diphoton ($\Mgg$), dijet ($\Mjj$),
and the four-body systems ($\Mggjj$), as well as the direction of Higgs bosons
in the Collins--Soper frame~\cite{Collins:1977iv}, to provide discrimination
between production of two Higgs bosons and SM background.

A search in the same final state was performed by the
ATLAS collaboration~\cite{Aad:2014yja}. Complementary final states such as $\HH \to \bbbar\bbbar$, $\HH \to \tau\tau\bbbar$, and $\HH$ to multileptons and multiphotons were also explored by the ATLAS \cite{Aad:2015uka, Aad:2015xja} and CMS~\cite{Khachatryan:2014jya, Khachatryan:2015year, Khachatryan:2016cfa, Khachatryan:2015tha} collaborations.

This paper is organized as follows: Section \ref{sec-detector} contains a brief description of the CMS detector.
In Section \ref{sec:data_sim} we describe the simulated signal and background event samples used in the analysis.
Section \ref{section:reconstruction} is dedicated to the discussion of event selection and Higgs boson reconstruction.
The signal extraction procedure is discussed in Section \ref{sec::AnalysisMethods}. In Section  \ref{section:sys} we present the systematic uncertainties
impacting each analysis method. Section \ref{section:results} contains the results of resonant and nonresonant searches, and Section 8 provides a summary.

\section{The CMS detector} \label{sec-detector}

The CMS detector, its coordinate system, and main kinematic variables used in the analysis are described in detail
in Ref.~\cite{Chatrchyan:2008zzk}. The detector is a multipurpose apparatus
designed to study physics processes at large transverse momentum
\pt in pp and heavy-ion collisions.
The central feature of the apparatus is a superconducting solenoid, of 6\unit{m} internal diameter, providing a
magnetic field of 3.8\unit{T}. A silicon pixel and strip tracker covering the pseudorapidity range $\abs{\eta}< 2.5$, a crystal
electromagnetic calorimeter (ECAL), and a brass and scintillator hadron calorimeter (HCAL) reside within the field volume.
The ECAL is made of lead tungstate crystals, while the HCAL has layers of plates
of brass and plastic scintillator. These calorimeters are both composed of a barrel and two endcap sections and provide coverage  up to $\abs{\eta}< 3.0$. An iron and quartz-fibre Cherenkov hadron calorimeter covers larger values of $3.0 < \abs{\eta}< 5.0$.
Muons are measured in the $\abs{\eta}< 2.4$ range, using
detection planes based on three technologies: drift tubes, cathode strip chambers, and resistive-plate chambers.

The first level of the CMS trigger system, composed of special hardware processors, uses information from the calorimeters and muon detectors to select the most interesting events in a time interval of less than 4\mus. The high-level trigger (HLT) processor farm further decreases the event rate from around 100\unit{kHz} to less than 1\unit{kHz}, before data storage.

\section{Simulated events} \label{sec:data_sim}

The \MADGRAPH version 5.1.4.5~\cite{Alwall:2014hca} Monte Carlo (MC) program generates
parton-level signal events based on matrix element calculations at leading order (LO) in quantum chromodynamics (QCD),
using LO \PYTHIA version 6.426~\cite{Pythia6-0}  for showering and hadronization of partons.
The models provide a description of production through gluon-gluon fusion of particles with narrow width (width set to 1\MeV) that decay to two Higgs bosons, with mass $\mH = 125\GeV$, in agreement with Ref. \cite{Aad:2015zhl}. Events are generated either for spin-0 radion production, or spin-2 KK-graviton production predicted by the bulk-RS model.

The samples for nonresonant production are generated considering the cross section dependence on three parameters: the Higgs boson trilinear coupling $\lambda$,
parametrized as $\kapl \equiv \lambda/\lbdSM$, where $\lbdSM \equiv \mH^2 /(2 v^2) = 0.129$, with $v = 246\GeV$ being the vacuum expectation value of the Higgs boson; the top Yukawa coupling $\yt$, parametrized
as $\kapt \equiv \yt/\yt^\mathrm{SM}$, where $\yt^\mathrm{SM} = \mt/v$ is the SM value of the top Yukawa coupling, and $\mt$ the top quark mass; and the coefficient $\ctwo$ of a possible coupling of two Higgs
bosons to two top quarks. The first two parameters reflect changes relative to SM values, while the third corresponds purely to a BSM operator.
In this parametrisation the SM production corresponds to the point $\kapl = 1$, $\kapt = 1$ and $\ctwo = 0$.
The parameters $\kapl$ and $\ctwo$ cannot be directly constrained by alternative measurements at the LHC. Therefore we vary these parameters in a wide range: $-20 \leq \kapl \leq 20$ and $-3 \leq \ctwo \leq 3$.
The range $ 0.75 \leq \kapt \leq 1.25$ is compatible with constraints from the single Higgs boson measurements provided in Ref.~\cite{Khachatryan:2014jba}.

The part of the Higgs potential $\Delta \mathcal{L}$ relevant to two-Higgs boson production and their interactions with the top quark can be expressed
as in Ref.~\cite{Corbett:2015mqf}:
\begin{equation}
\Delta \mathcal{L} =
\kapl\, \lbdSM v\, H^3
- \frac{\mt}{v}(v+  \kapt\,   H  +  \frac{c_2}{v} HH ) \,( {\overline{t}_\mathrm{L}}t_\mathrm{R} + \text{h.c.}),
\label{eq:Hpot}
\end{equation}
where  $t_\mathrm{L}$ and $t_\mathrm{R}$ are the top quark fields with left and right chiralities, respectively, and $H$ is the physical Higgs boson field.

Besides being used to predict SM single-Higgs boson production, the MC predictions for the background processes
are used also in comparisons with data, to optimize the selection criteria, and for checking background-estimation methods based on control samples in data.
The dominant background, originating from events with two prompt photons and two jets in the final state, is generated at next-to-leading-order (NLO) in QCD  using \SHERPA version 1.4.2~\cite{Gleisberg:2008ta}.
Multijet production with or without a single-prompt
photon represents a subdominant background, and is generated with the \PYTHIA6 package.
Other minor backgrounds, including Drell--Yan ($\Pp\Pp \to \Z/\gamma^* \to \Pep\Pem$),
SM Higgs boson production with jets, as well as vector boson and top quark production in association with photons, are generated
using \MADGRAPH and \PYTHIA6, or the generator \POWHEG version 1.0~\cite{Nason:2004rx, POWHEG_Frixione:2007vw, Alioli:2010xd} at NLO in QCD.
The generated events are processed through \GEANTfour-based~\cite{Agostinelli:2002hh,GEANT}
detector simulation.

\section{Event reconstruction}
\label{section:reconstruction}

The events are selected using two complementary HLT paths requiring two photons. The first trigger requires an identification
based on the energy distribution of the electromagnetic
shower and loose isolation requirements on photon candidates.
The second trigger applies tighter constraints on the shower shape, but a looser kinematic selection.
The trigger thresholds on the \pt range between 26 and 36\GeV, and between 18 and 22\GeV, respectively, for photons with highest (leading) and next-to-highest (subleading) \pt, with specific choices that depend on the instantaneous LHC luminosity.
The HLT paths are more than 99\% efficient for the selection criteria used in this analysis~\cite{Khachatryan:2014ira}.

\subsection{The \texorpdfstring{$\PH\to \gamma \gamma$}{Higgs to two photons} candidate}

Photon candidates are constructed from clusters of energy in the ECAL~\cite{CMS-PAS-EGM-10-005, CMS-PAS-EGM-10-006}.
They are subsequently calibrated \cite{ECALpaper} and identified through a cutoff-based approach (referred to as ``cut-based analysis" in Ref.~\cite{Khachatryan:2014ira}).
The identification criteria include requirements on \pt of the electromagnetic shower, its longitudinal leakage into the HCAL, its isolation from jet activity in the event, as well as a veto on the presence of a track matching the ECAL cluster. These criteria provide efficient rejection of objects that arise from jets or electrons but are reconstructed as photons.
Both photons are required to be within the ECAL fiducial volume of $\abs{\eta_{\gamma}}<2.5$.
Small transition regions between the ECAL barrel and the ECAL endcaps are excluded in this analysis, because the reconstruction of a photon object in this region is not optimal.

The directions of the photons are reconstructed assuming that they arise from the primary vertex of the hard interaction.
However on average $\approx$20 additional pp interactions (pileup) occur in the same or neighboring pp bunch crossings as the main interaction.
Many additional vertexes are therefore usually reconstructed in an event using charged particle tracks.
We assume that the primary interaction vertex corresponds to the one that maximizes the sum in $\pt^2$ of the associated charged particle tracks.
For the simulated signal, it is shown that this choice of vertex lies within 1\cm
of the true hard-interaction vertex in 99\% of the events.
With this choice for energy reconstruction and vertex identification, the diphoton mass resolution remains close to 1\GeV
independent of the signal hypothesis.

Diphoton candidates are preselected by requiring $100 < \Mgg < 180\GeV$.
The two photons are further required to satisfy the asymmetric selection criteria
$\pTgone/ \Mgg > 1/3$ and
$\pTgtwo/ \Mgg > 1/4$,
where $\pTgone$ and $\pTgtwo$ are the transverse momenta
of the leading and subleading photons.
The use of different \pt thresholds scaled by the diphoton invariant mass,
minimizes turn-on effects that can distort the distribution at the low-mass end of
the $\Mgg$ spectrum.
If there is more than one diphoton candidate selected through the above requirements,
the pair with the largest scalar sum in the \pt of the two photons is chosen for analysis.

\subsection{The \texorpdfstring{$\PH\to \bbbar$}{Higgs to two \PQb quarks} candidate}

The Higgs boson candidate decaying into two \PQb quarks is reconstructed
following a procedure similar to that used in CMS searches for
SM Higgs bosons that decay to \PQb quarks~\cite{HbbLegacy}.

The particle-flow event algorithm reconstructs and identifies each individual particle (referred to as candidates) with an optimized combination of information from the various elements of the CMS detector~\cite{PFPAS2009, CMS-PAS-PFT-10-001}.
Then, the anti-\kt algorithm~\cite{AK5} clusters particle-flow candidates into jets using a distance
parameter $D = 0.5$.
Jets are required to be within the tracker acceptance ($\abs{\etaj} < 2.4$), and separated from both photons through a condition on the angular distance in $\eta{\times}\phi$ space of $\DRgj \equiv \sqrt{\smash[b]{(\Delta \eta)^2 +(\Delta \phi)^2}}  > 0.5$, where $\phi$ is the azimuth angle in radians.
The jet energy is corrected for extra depositions from pileup interactions,
using the jet-area technique~\cite{Cacciari:2007fd} implemented in the {\sc FastJet}
package~\cite{Cacciari:2011ma}.
Jet energy corrections are applied as a function of $\etaj$ and $\pTj$~\cite{JINST6, CMS-DP-2013-011}.
Identification criteria are applied to reject detector noise misidentified as jets, and the procedure
is verified using simulated signal.

The identification of jets likely to have originated from hadronization of \PQb quarks exploits
the combined secondary vertex (CSV) \PQb quark tagger~\cite{BTV}. This algorithm combines the
information from track impact parameters and secondary vertexes within a given jet into a continuous output discriminant.
Jets with CSV tagger values above some fixed threshold are considered as \PQb tagged.
The working point chosen in this analysis corresponds to an efficiency, estimated from simulated multijet events, of $\approx$70\% and a mistag rate for light quarks and gluons of
1--2\%, depending on jet \pt.
This efficiency and the mistag rate are measured in data samples enriched in \PQb jets (\eg, in~$\ttbar$ events).
Correction factors of $\approx 0.95$ are determined from data-to-simulation comparisons and applied as weights to all simulated events.

Events are kept if at least two jets are selected and at least one of them is \PQb tagged.
To improve signal sensitivity, events are subsequently classified in two categories:
events with exactly one \PQb-tagged jet (medium purity) and events with more than one \PQb-tagged jet (high purity).
In the former category, the $\PH\to \bbbar$ decay is reconstructed by pairing the \PQb-tagged jet with a non \PQb-tagged jet,
while in the latter category a pair of \PQb-tagged jets is used.
In both cases, when multiple pairing possibilities exist for the Higgs boson candidate, the dijet system with largest \pt is retained for further study.
For medium- and high-purity simulated signal events, this procedure selects the correct jets in more than 80\% and more than 95\%, respectively.

The resolution in $\Mjj$ improves from 20\GeV for $\mx = 300\GeV$ to 15\GeV for $\mx = 1\TeV$ in the
high-purity category, and
from 25\GeV for $\mx = 300\GeV$ to 15\GeV for $\mx = 1\TeV$
in the medium-purity category.
In the search for a low-mass resonance, the dijet mass resolution is improved
using a multivariate regression technique~\cite{HbbLegacy}
that  uses the global information from the events as well as the particular properties of each
jet, in an attempt to identify the semi-leptonic decays of \PB~mesons and correct for the energy carried away by undetected neutrinos.
The relative improvement in resolution
is typically 15\%.
For the high mass analysis and nonresonant analysis the $\Mjj$ resolution is better than for low mass analysis.
The improvement provided by the regression technique was found to be very limited. Therefore in those cases no regression was used.

Independent of whether a search involves the usage of jet energy regression, all jets are
required to have $\pTj > 25\GeV$. Finally, we require that $ 60 < \Mjj < 180\GeV$.

\subsection{The two-Higgs-boson system}

The object selections discussed thus far are summarized in Table~\ref{table:gencut}.
\begin{table}[htpb]
\topcaption{Summary of the analysis preselections.}
\label{table:gencut}
\centering
\renewcommand\arraystretch{1.3}
\begin{scotch}{lc{c}@{\hspace*{5pt}}lc}
\multicolumn{2}{c}{Photons} && \multicolumn{2}{c}{Jets} \\
\cline{1-2}\cline{4-5}
Variable & Range && Variable & Range \\[\cmsTabSkip]
$\pTgone / \Mgg$ & ${>}1/3$  &&   $\pTj$ (\GeVns)& $>$25              \\
$\pTgtwo / \Mgg$ & ${>}1/4$  && &                             \\
$\abs{\eta_{\gamma}}$ & $<$2.5        && $\abs{\etaj}$ & $<$2.4              \\
$ \Mgg$  (\GeVns) & $[100, 180]$   && $\Mjj$ (\GeVns) & $[60,180]$      \\
                   &        &&  \PQb-tagged jets & $>$0    \\
\end{scotch}
\end{table}

In each category, two Higgs bosons are obtained by combining the diphoton and the dijet boson candidates.
To improve the resolution in $\Mggjj$, an
additional constraint is imposed requiring $\Mjj$ to be
consistent with $\mH$. This is achieved by modifying the jet 4-momenta using multiplicative factors.
The value of each factor is obtained event-by-event through a $\chi^2$ minimization procedure where the size of the denominator is defined by the estimated resolution for each jet~\cite{Chatrchyan:2013yoa}.
The procedure, similar to the one used  in Ref.~\cite{Chatrchyan:2013yoa}, is referred to as a kinematic fit and the resulting four-body mass is termed $\Mggjjk$.

The scattering angle, $\thetastar$, is defined in the Collins--Soper frame of the four-body system state,
as the angle between the momentum of the Higgs boson decaying into two photons
and the line that bisects the acute angle between the colliding protons.
In the Collins--Soper frame, the two Higgs boson candidates are collinear,
and the choice of the one decaying to photons as reference is therefore arbitrary.
Using the absolute value of the cosine of this angle, $\acosthetastar$,
obviates this arbitrariness.

\subsection{Backgrounds}

The SM background in $\Mgg$ can be classified into two categories: the nonresonant background, from multijet and electroweak processes, and a peaking background corresponding to events from single-Higgs bosons decaying to two photons.

After the baseline selections of Table \ref{table:gencut}, the dominant nonresonant background with two prompt photons and more than two extra jets, referred to as $\gamma\gamma + {\geq}2\,\text{jets}$, represents $\approx$75\%
of the total background. The nonresonant background with one prompt photon and a jet
misidentified as a photon as well as more than two extra jets, referred to as $\gamma\,\text{jet}+ \geq2\,\text{jets}$, represents in turn $\approx 25\%$.
The background from two jets misidentified as photons is negligible.

The remaining nonresonant and resonant backgrounds
contribute much less than 1\% to the total. They represent associated production of photons with top quarks or single
electroweak bosons decaying to quarks,
and Drell--Yan events with their decay electrons misidentified as photons.
The resonant backgrounds correspond to different
SM processes contributing to single-Higgs boson production.

All nonresonant backgrounds are estimated from data, and the resonant background from SM single-Higgs boson production in different channels is taken from the MC
simulation normalized to NLO or next-to-NLO (NNLO) production cross sections, whichever are available \cite{Heinemeyer:2013tqa}.

\begin{figure*}[hbt]\centering
\includegraphics[width=0.49\textwidth]{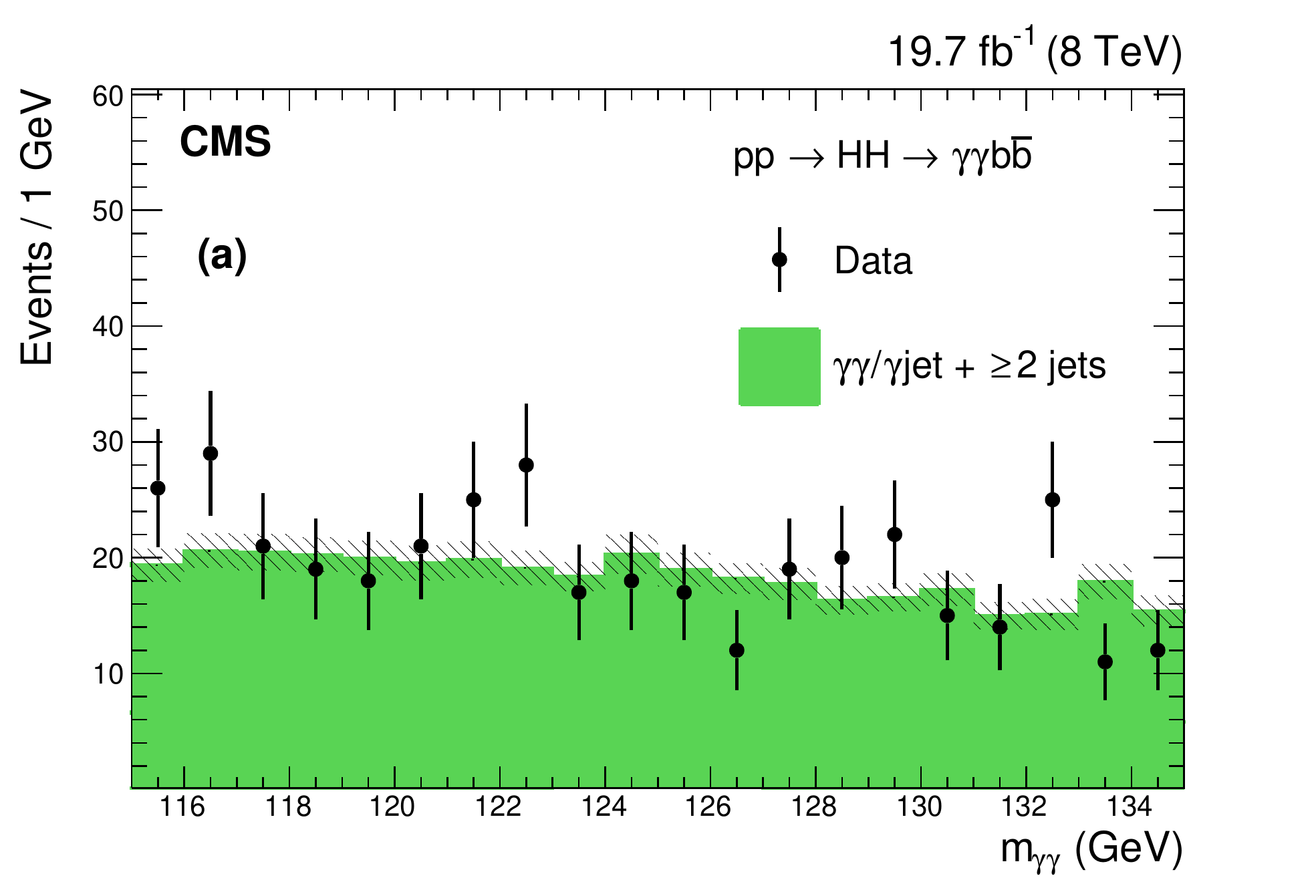}
\includegraphics[width=0.49\textwidth]{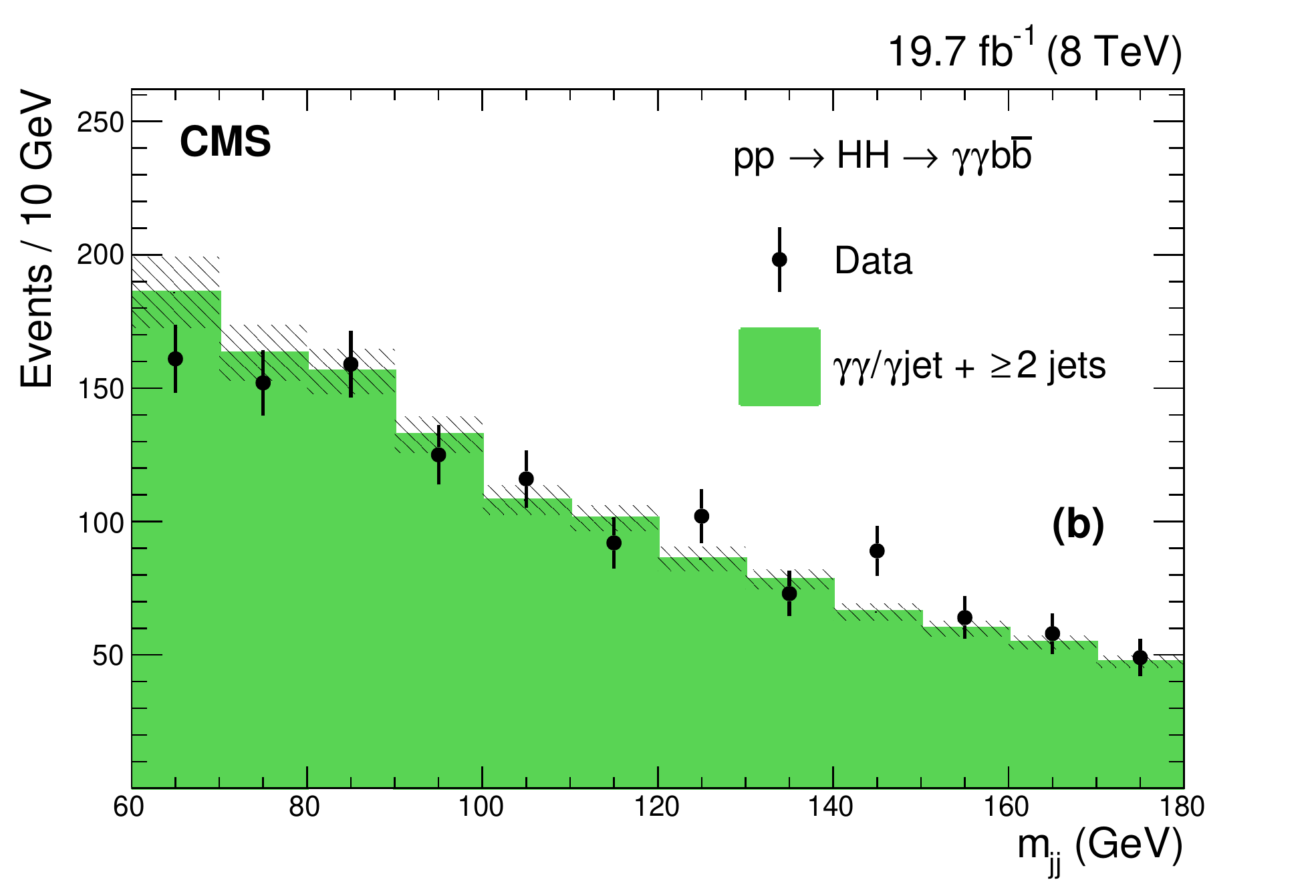}
\includegraphics[width=0.49\textwidth]{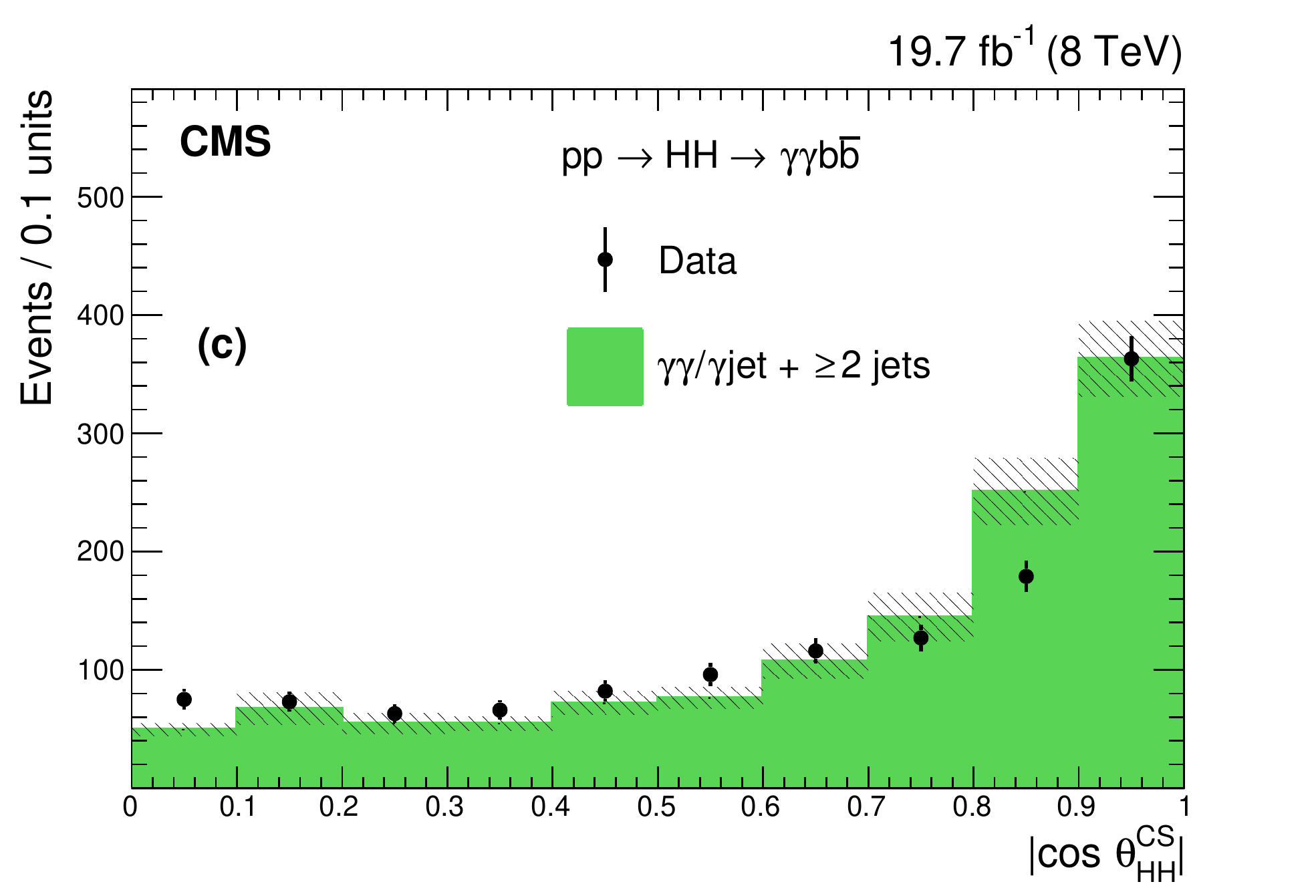}
\includegraphics[width=0.49\textwidth]{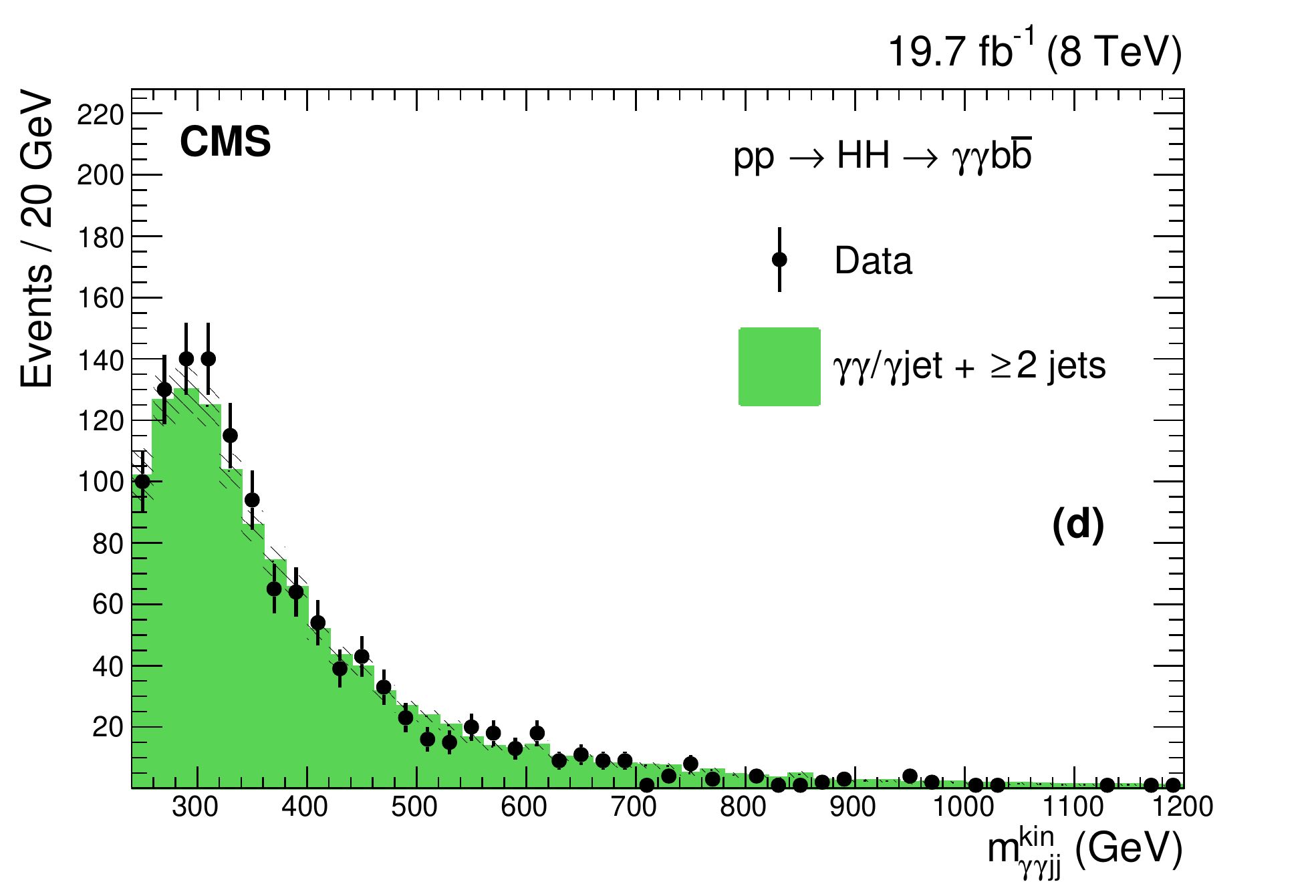}
\caption{
Reconstructed spectra for data compared to the $\gamma\gamma/\gamma\,\text{jet} + {\geq}2\,\text{jets}$ background after the selections described in Table \ref{table:gencut} (selections on photons and jets and a requirement of at least one \PQb-tagged jet):
(a)~$\Mgg$, (b)~$\Mjj$, (c)~$\acosthetastar$, and (d)~$\Mggjjk$.
The hatched area corresponds to the bin-by-bin statistical uncertainties on the background prediction reflecting the limited size of the generated
MC sample.
The comparison is provided for illustrative purpose,
in the backgrounds, except the one coming from single-Higgs production, are evaluated from a fit to the data without reference to the MC simulation.}
\label{figure:control_plots}\end{figure*}

The comparison between data and MC predictions is provided in Fig.~\ref{figure:control_plots}.
The $\gamma\gamma/\gamma\,\text{jet}+ {\geq}2\,\text{jets}$ background is normalized to the total integral of data in the signal free region,
defined by the condition $\Mgg > 130$ or $\Mgg < 120\GeV$ in addition to the selections of Table \ref{table:gencut}.

\section{Analysis methods}
\label{sec::AnalysisMethods}

\begin{figure*}[hbt]\centering
\includegraphics[width=0.49\textwidth]{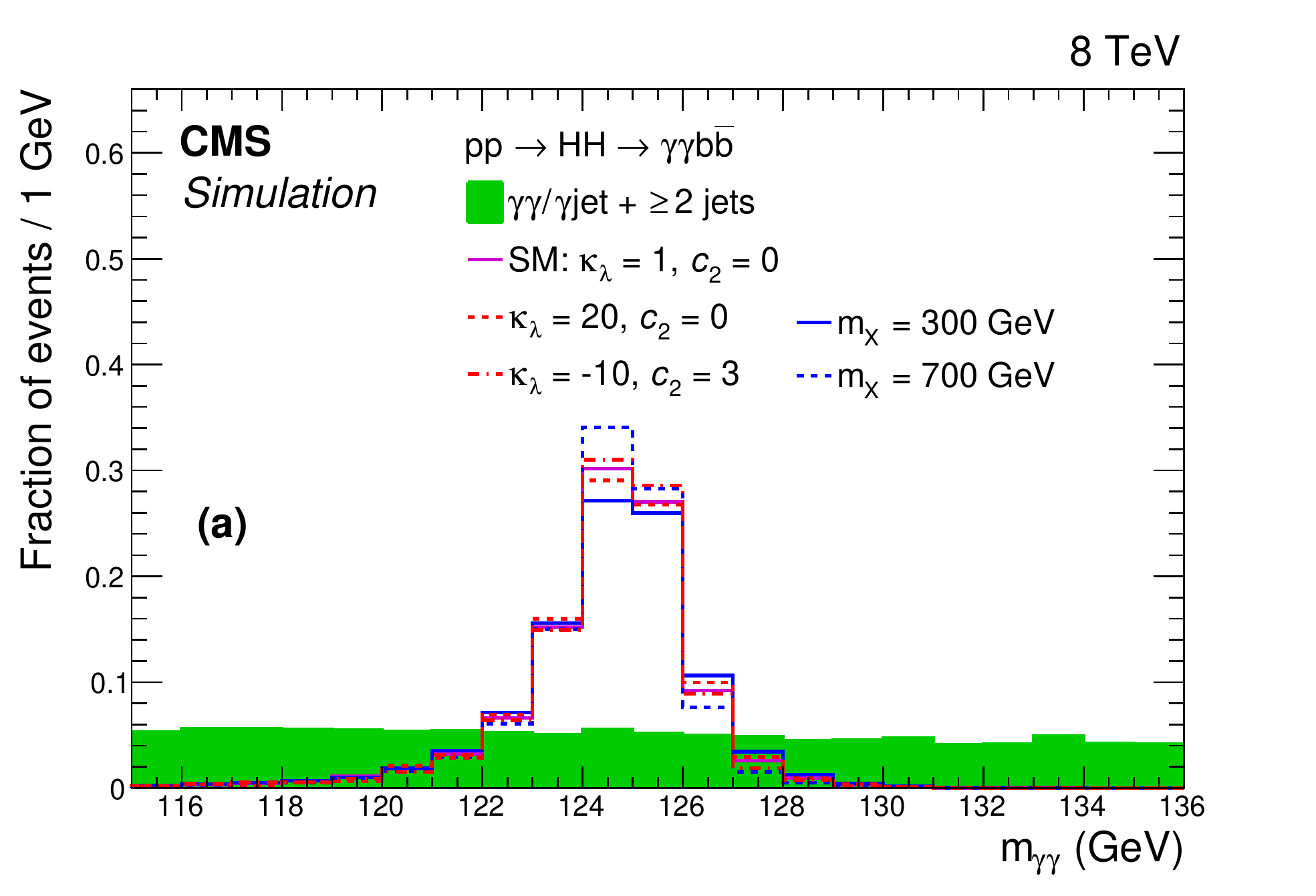}
\includegraphics[width=0.49\textwidth]{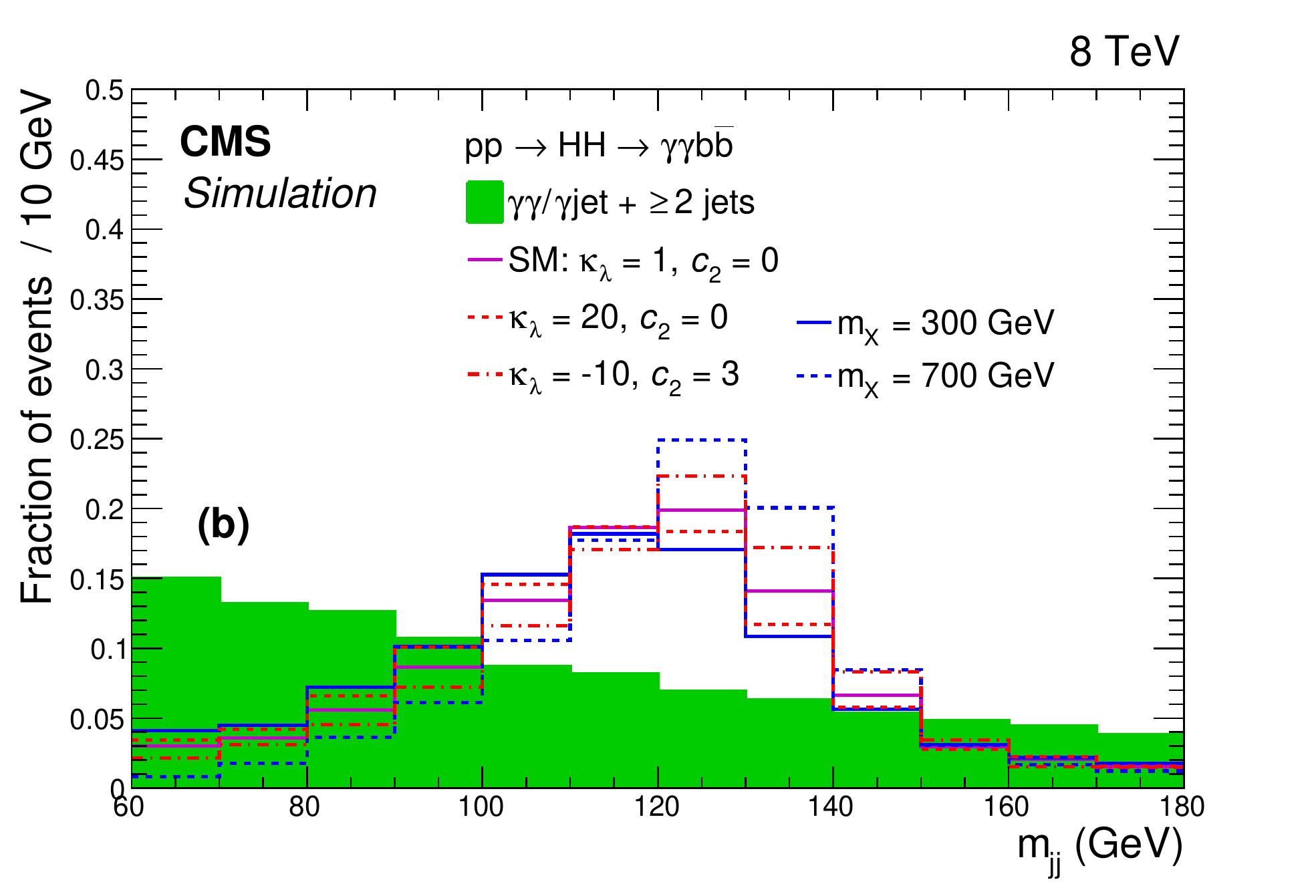}
\includegraphics[width=0.49\textwidth]{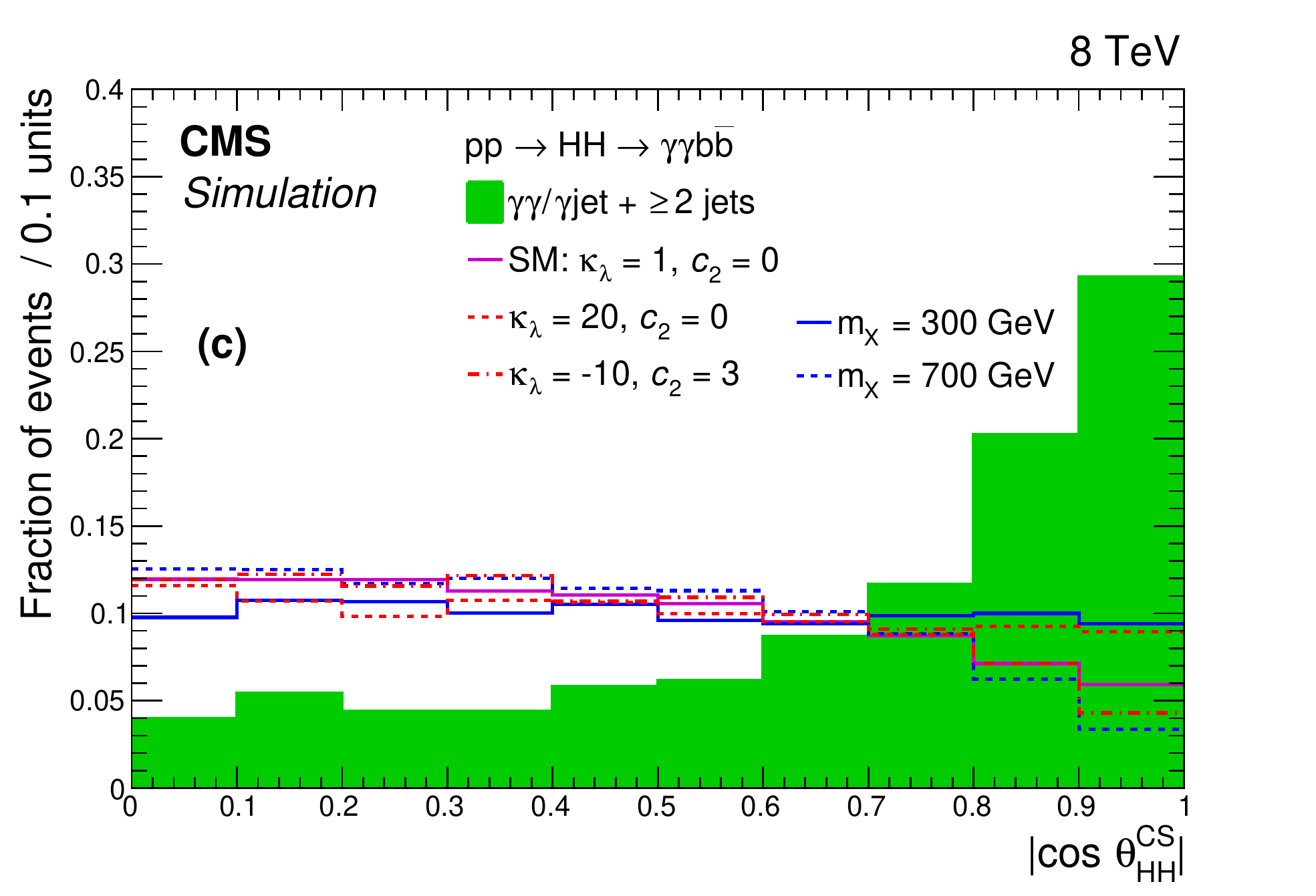}
\includegraphics[width=0.49\textwidth]{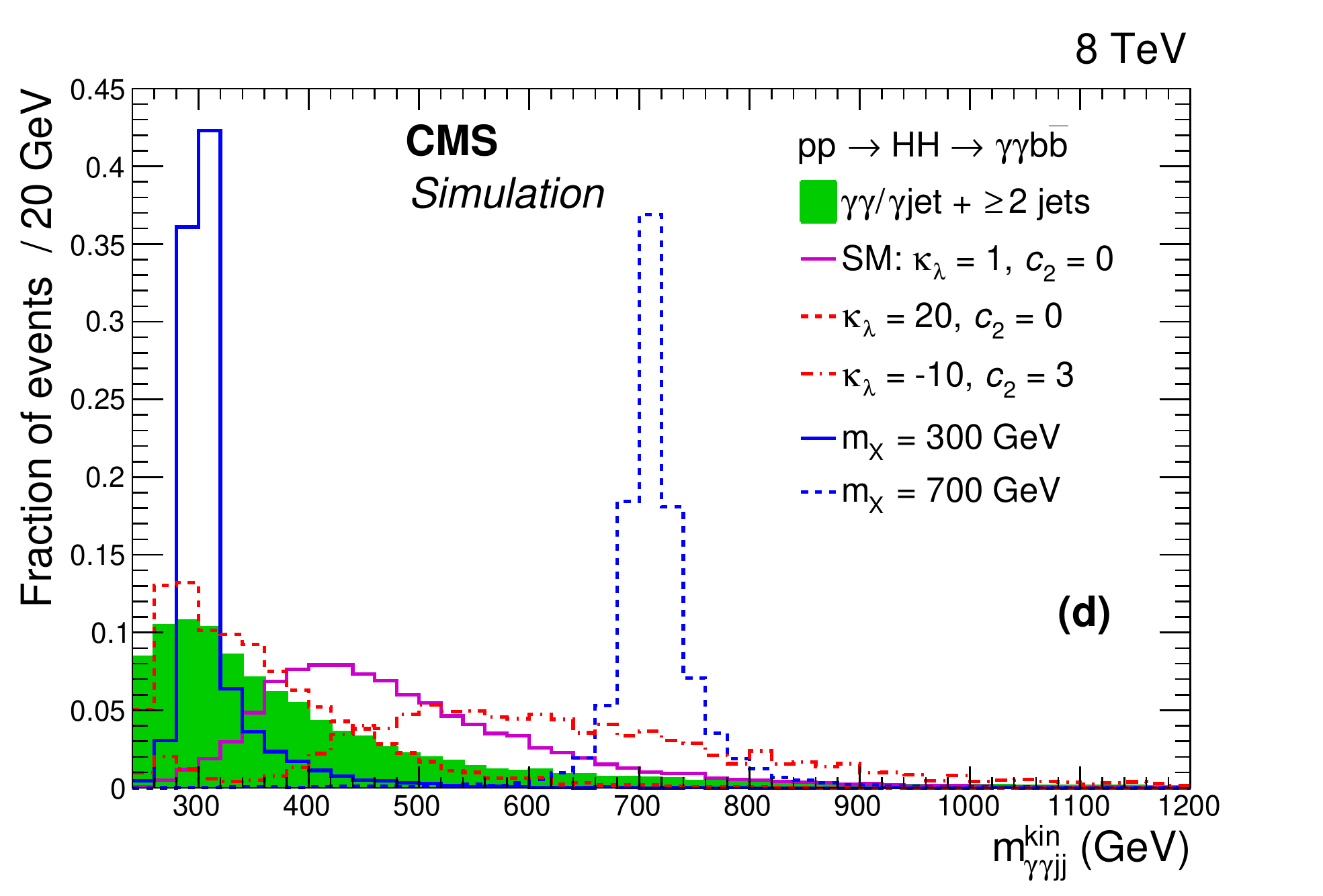}
\caption{Simulated spectra for the spin-0 radion signal at $\mx=300$ and 700\GeV, and for some values of the anomalous couplings, compared to SM Higgs boson production and QCD background, after the selections described in Table \ref{table:gencut} (selections on photons and jets and a requirement of at least one \PQb-tagged jet): (a) $\Mgg$, (b) $\Mjj$, (c) $\acosthetastar$, and (d) $\Mggjjk$. All spectra are normalized to unity.}
\label{figure:massea_signal}\end{figure*}

In the final step, this analysis exploits kinematic properties
of the final state to discriminate either resonant or nonresonant signal from SM background:
the Higgs boson masses $\Mgg$ and $\Mjj$, the cosine of their
scattering angle $\acosthetastar$, and the mass of the two-Higgs-boson system, $\Mggjjk$. Distributions in these variables are shown for different signal assumptions in Fig.~\ref{figure:massea_signal}. The signal peaks in $\Mgg$ and $\Mjj$ are shown in Figs.~\ref{figure:massea_signal}(a) and (b). The corresponding distributions for the QCD background are smoothly varying over the shown ranges.
The $\acosthetastar$ is rather uniform for signal, as shown in Fig.~\ref{figure:massea_signal}(c), while it peaks toward one for background.
Finally, a resonant signal appears as a narrow peak in $\Mggjjk$ spectrum, while the nonresonant signal has a broad contribution as shown in Fig.~\ref{figure:massea_signal}(d).

The dominant background from non-resonant production of prompt photons and jets exhibits a kinematic peak around $\Mggjjk \approx 300\GeV$ followed by a slowly falling tail at high $\Mggjjk$. In the resonant case, we consider two strategies, one for $\mx$ close to the kinematic peak, and one for $\mx$ heavier than the kinematic peak. A third strategy is considered for the nonresonant case, since the signal distribution as function of $\Mggjjk$ is broad.
In all cases a categorization is used based on the number of \PQb-tagged jets.
All the strategies are summarized in Table~\ref{table:methods} and briefly described below:

\begin{enumerate}

\item {Resonant search in the low-mass region} ($260 \leq \mx \leq 400\GeV$):
the events are selected in a narrow window around the $\mx$
hypothesis in the $\Mggjjk$ spectrum, and the signal is identified simultaneously in the $\Mgg$ and $\Mjj$ spectra. This approach avoids a direct search for a resonance in the $\Mggjjk$ spectrum near the top of the kinematic peak of the SM background.

\item {Resonant search in the high-mass region} ($400 \leq \mx \leq 1100\GeV$):
the events are selected in a window around $\mH$ in both the $\Mgg$ and $\Mjj$ spectra, and
the signal is identified in the $\Mggjjk$ spectrum.

\item {Nonresonant search}: a selection is applied in the $\acosthetastar$ variable to reduce the background.
In addition to the categorization in the number of \PQb-tagged jets,
a categorization is applied in $\Mggjjk$ by defining a high-mass region and a low-mass region.
The signal is identified simultaneously in the $\Mgg$ and $\Mjj$ spectra.

\end{enumerate}

\begin{table*}[htb]
\renewcommand{\arraystretch}{1.3}
\topcaption{Summary of the search analysis methods.
\label{table:methods}}
\centering
\begin{scotch}{   llll }
Signal hypothesis  & Select & \# categories & Fit \\\hline
(1) $\mx \le $ 400\GeV  & $\Mggjjk$ & 2 (\PQb tags) & $\Mgg$, $\Mjj$ \\
(2) $\mx \ge$ 400\GeV  & $\Mgg$, $\Mjj$  & 2 (\PQb tags) & $\Mggjjk$  \\
(3) Nonresonant  & $\acosthetastar$  &  4 (\PQb tags, $\Mggjjk$) &   $\Mgg$, $\Mjj$ \\
\end{scotch}
\end{table*}

The nonresonant background is described through different functions such as exponentials,
power-law, or polynomials in the Bernstein basis \cite{Khachatryan:2014ira}.
When the search is performed simultaneously
in the diphoton and dijet mass spectra,
these functions are used to construct a two-dimensional (2D)
probability density (PD) for the background in each category, following an approach similar to Ref.~\cite{Chatrchyan:2012vva}.
Otherwise, a one-dimensional (1D) PD is used.
In all cases, we choose the background PD to minimize the bias on signal.
The bias is always found to be at least a factor of 7 smaller than the statistical uncertainty in the
fit, and can be safely neglected~\cite{Chatrchyan:2012ufa}.

In each invariant mass distribution used to identify the signal, the signal PD
is modeled, following the same approach as in Ref.~\cite{Khachatryan:2014ira}, through the sum of a Gaussian function and a Crystal Ball (CB) function~\cite{CrystalBallRef}, using the parameters
extracted from fits to MC simulations.
The resolution parameters in both functions are kept independent,
$\sigma^\mathrm{G}_\mathrm{x}$ for the Gaussian and $\sigma^\mathrm{CB}_\mathrm{x}$ for the CB function, but in the fits to each of the channels ($\mathrm{x} = \gamma\gamma$, jj, $\gamma\gamma\mathrm{jj}$), we let
the $\mu$ parameter for both the Gaussian and the CB component float,
which  provides three independent $\mu_{\rm x}$ values.

Finally, we consider the contribution from SM single-Higgs boson production in 2D searches. The
gluon-gluon and vector-boson fusion processes are modeled in $\Mgg$ by a sum over
Gaussian and CB functions, and through a constant term in $\Mjj$. The
associated production of vector bosons that subsequently decay to jets, and the SM single-Higgs
bosons are modeled in the same way as the signal. The parameters of the distribution
are extracted from a fit to the MC simulation.

The total PD used for signal extraction corresponds to a sum over separate PD contributions from the signal component, single-Higgs boson production, and nonresonant backgrounds. We also verify that 2D PD functions can be considered as uncorrelated between $\Mgg$ and $\Mjj$ within the statistical uncertainties.
To obtain this result we calculated the correlation in data. The uncertainty in the correlation was estimated by generating pseudo-experiments from a model assuming no correlation between $\Mgg$ and $\Mjj$ and calculating the root mean square of the resulting distribution.

\subsection{Low-mass resonant}
\label{sec:LM}

In addition to the preselections summarized in Table~\ref{table:gencut}, each mass hypothesis has a selection
applied on $\Mggjjk$ in a narrow window around $\mx$. The window sizes increase with $\mx$ to account
for the increasing experimental resolution from $\Delta \mx = \pm 10$\GeV
at $\mx = 260$\GeV  to  $\Delta \mx =  ^{+31}_{-20}\GeV$ at $\mx = 400$\GeV.

\begin{figure*}[htb]\centering
\includegraphics[width=0.48\textwidth]{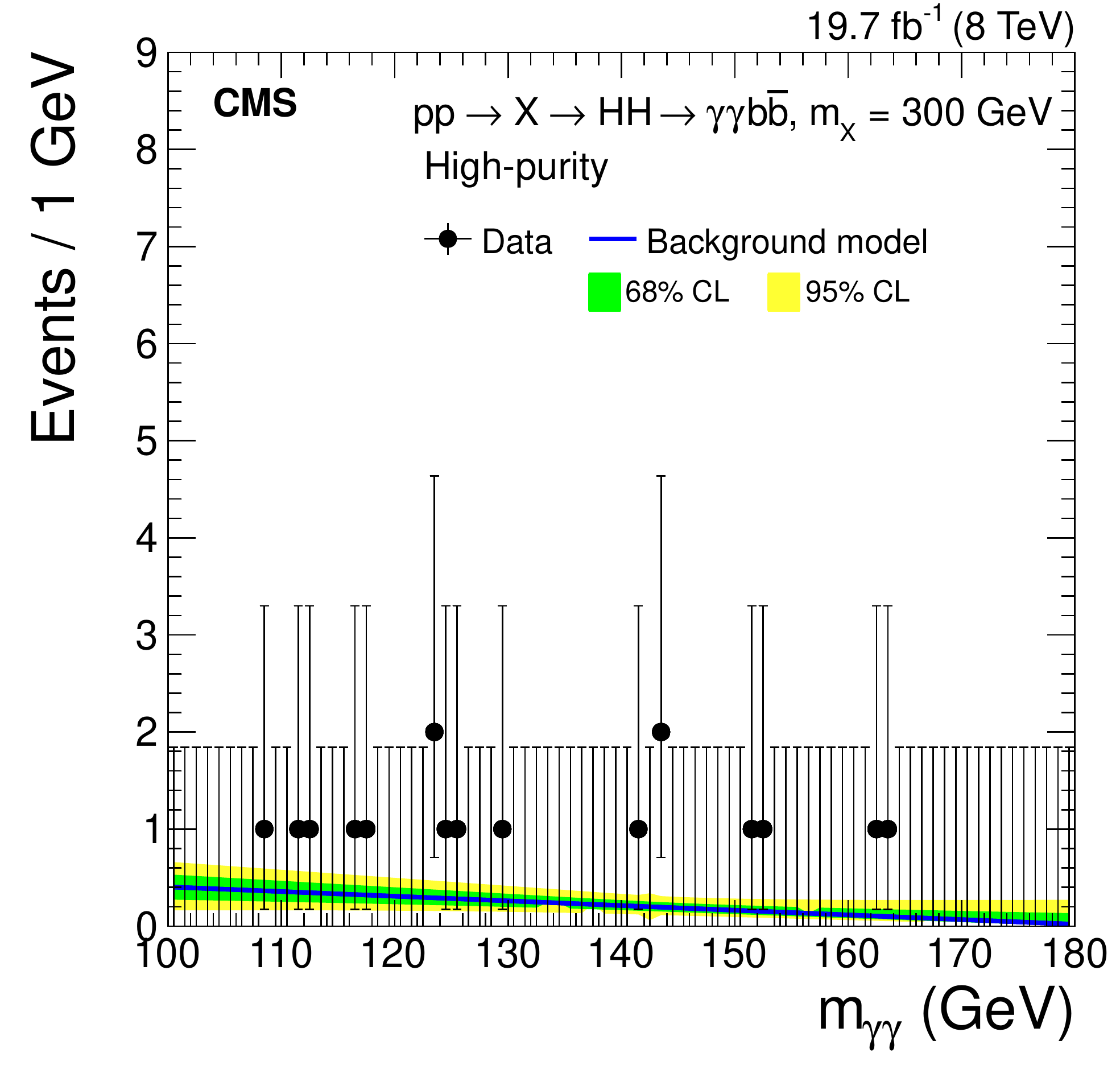}
\includegraphics[width=0.48\textwidth]{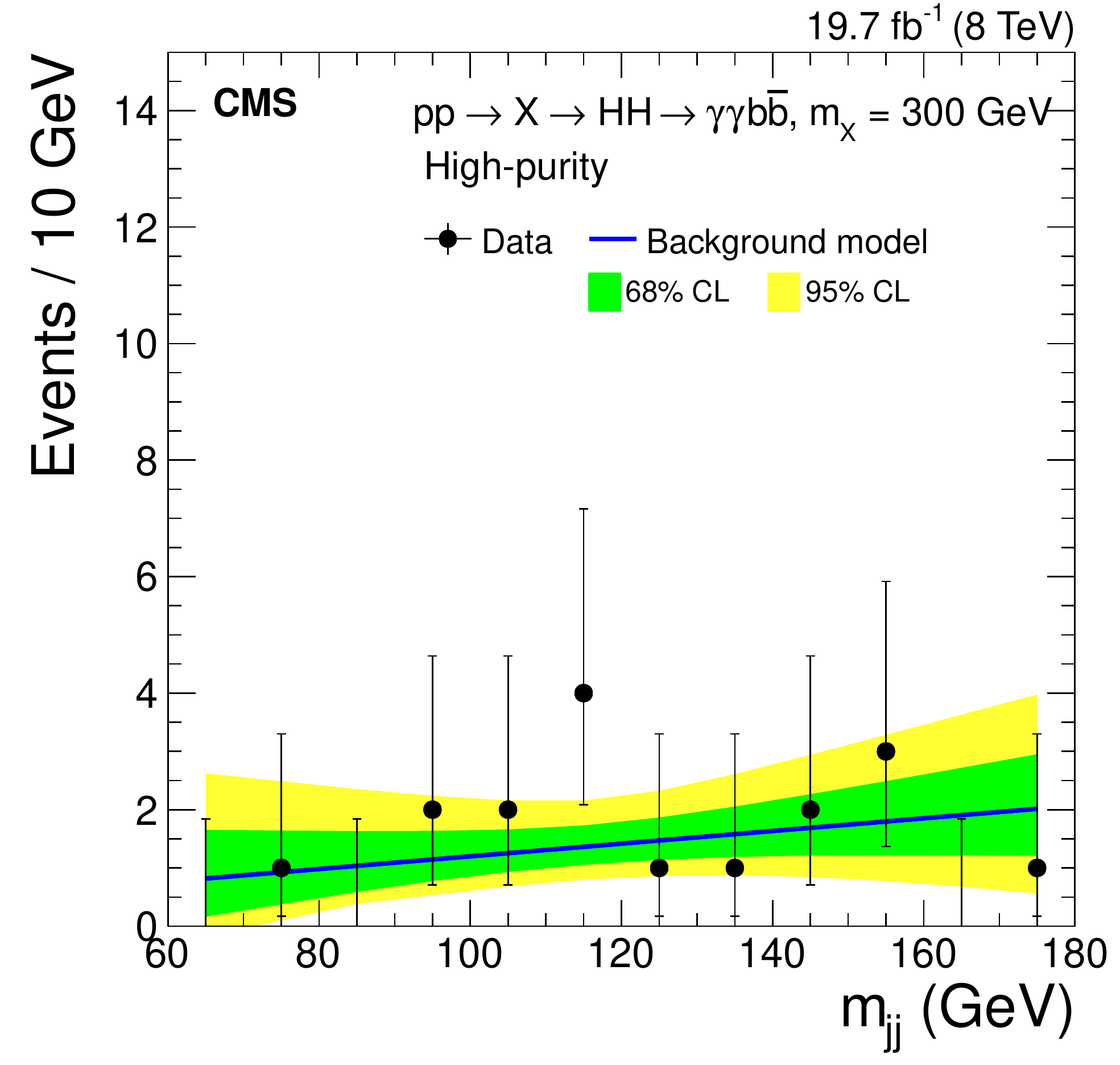}
\includegraphics[width=0.48\textwidth]{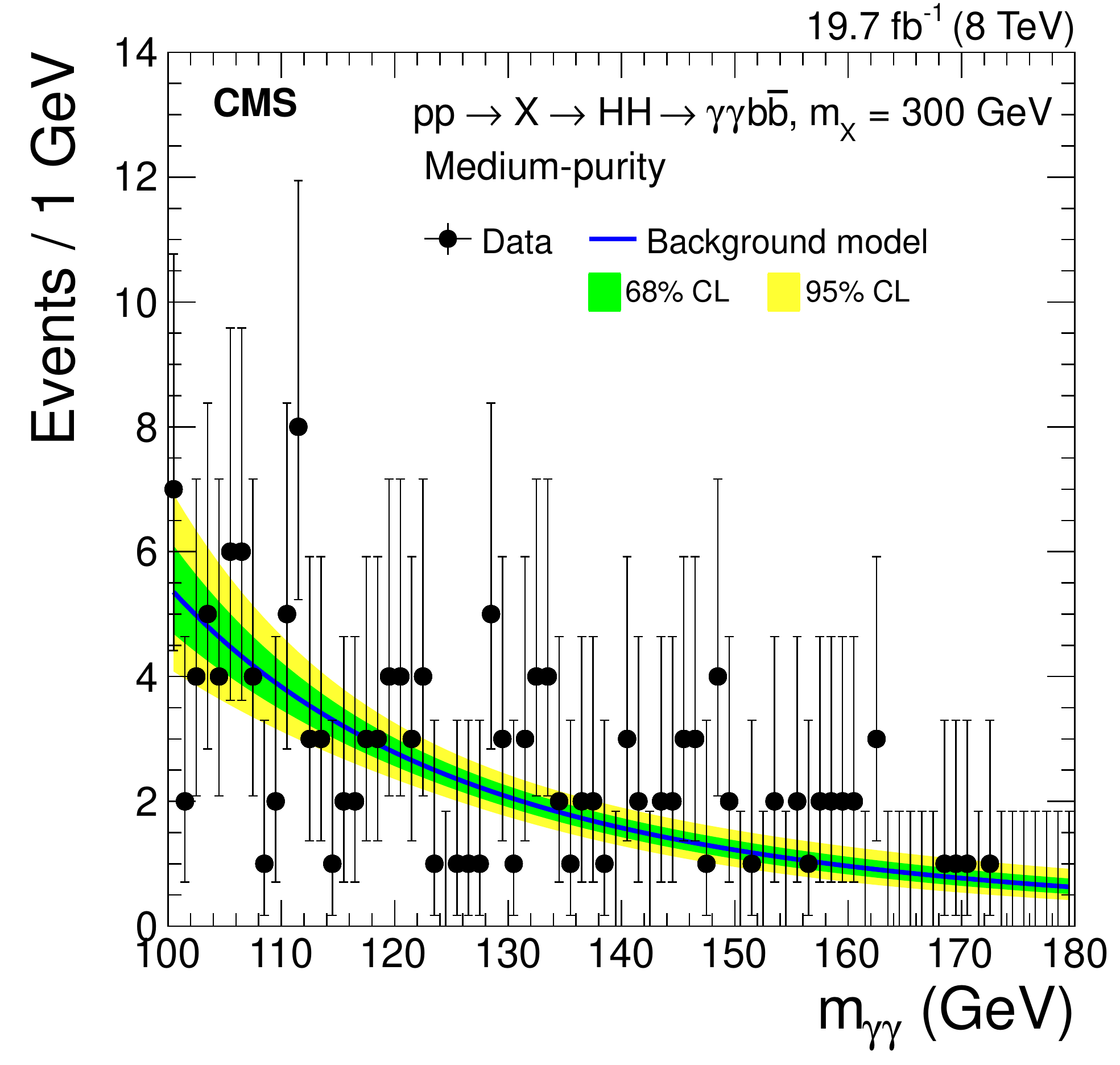}
\includegraphics[width=0.48\textwidth]{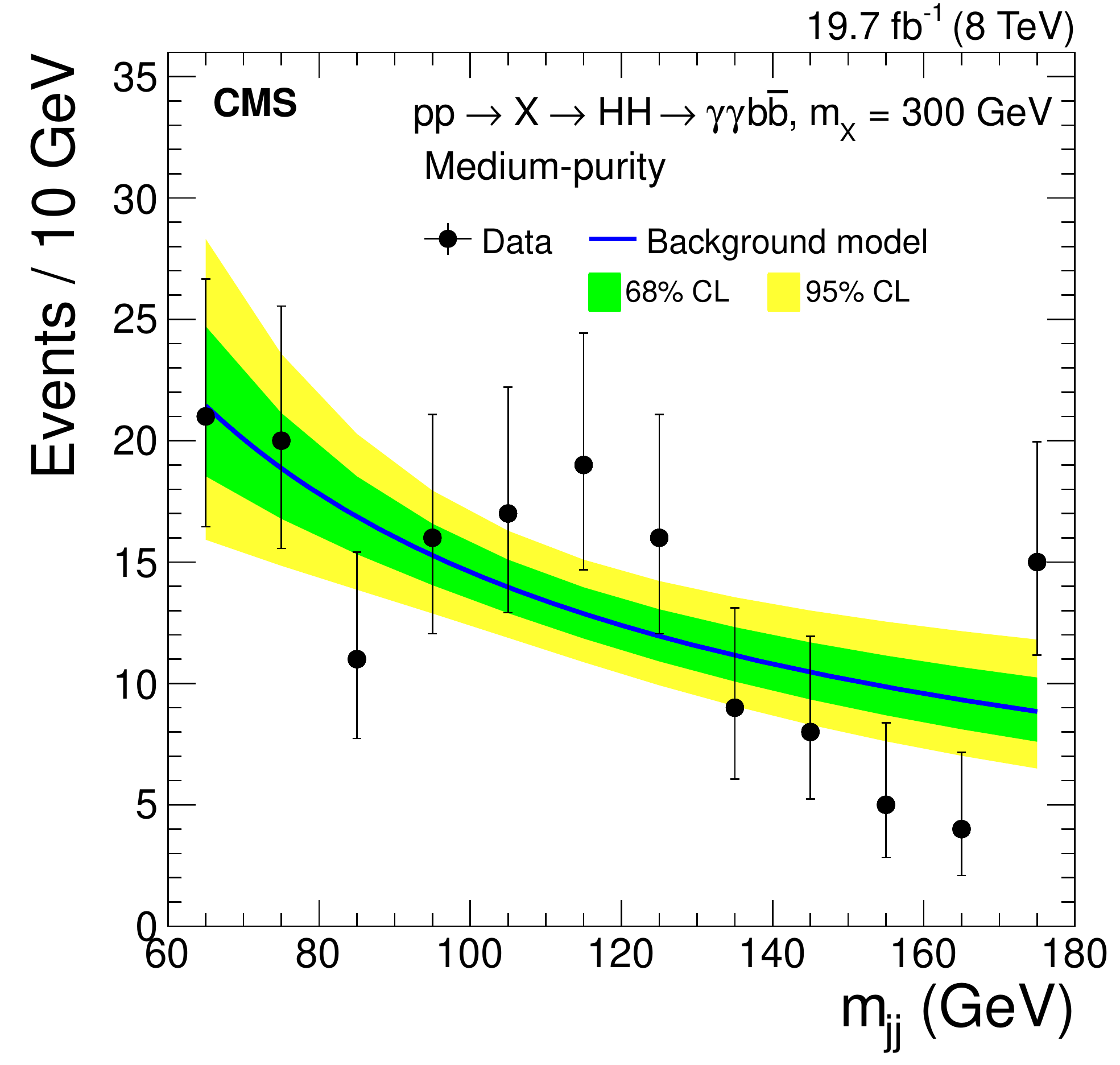}
\caption{Low-mass resonant analysis: fits to the nonresonant background contribution in high-purity category
 to the $\Mgg$ (top-left) and $\Mjj$ spectra (top-right),  and similarly
for medium-purity category in bottom-left and bottom-right, respectively.
The fits to the background-only hypothesis
are given by the blue curves, along with their 68\% and 95\% \CL contours. The selections are designed to search for a $\mx = 300\GeV$ hypothesis: $290 < \Mggjjk < 310\GeV$.}
\label{figure:BackgroundShapeMgg}\end{figure*}

A possible signal can be extracted from data using a simultaneous fit to the $\Mgg$ and
$\Mjj$ spectra. The sensitivity to the signal in this search is increased through the \PQb jet energy regression that improves the resolution of the signal in $\Mjj$. The background-only PD is a first-order polynomial in the Bernstein basis and a power law in the medium-
and high-purity categories, respectively, as shown in Fig.~\ref{figure:BackgroundShapeMgg}, together with their  68\% and 95\% confidence level (\CL) contours for the selection optimized for the search with $\mx = 300\GeV$,  $290 < \Mggjjk < 310\GeV$.

As a cross-check, two alternative signal extraction techniques are
tested.  In one, a selection is performed in the $\Mjj$ spectrum, and the signal extracted in the $\Mgg$
spectrum. In the other, a selection is performed in the $\Mjj$ spectrum
and the $\Mggjj$ spectrum is exploited, using a normalization extracted from
sidebands in the $\Mgg$ spectrum. The two procedures give
compatible results within the statistical uncertainties.

\subsection{High-mass resonant}
\label{sec:HM}

In addition to the requirements in Table~\ref{table:gencut},
selections are applied on $\Mgg$ and $\Mjj$, as summarized in
Table~\ref{table:mass_cuts_high}.

\begin{figure*}[htbp]\centering
\includegraphics[width=0.48\textwidth]{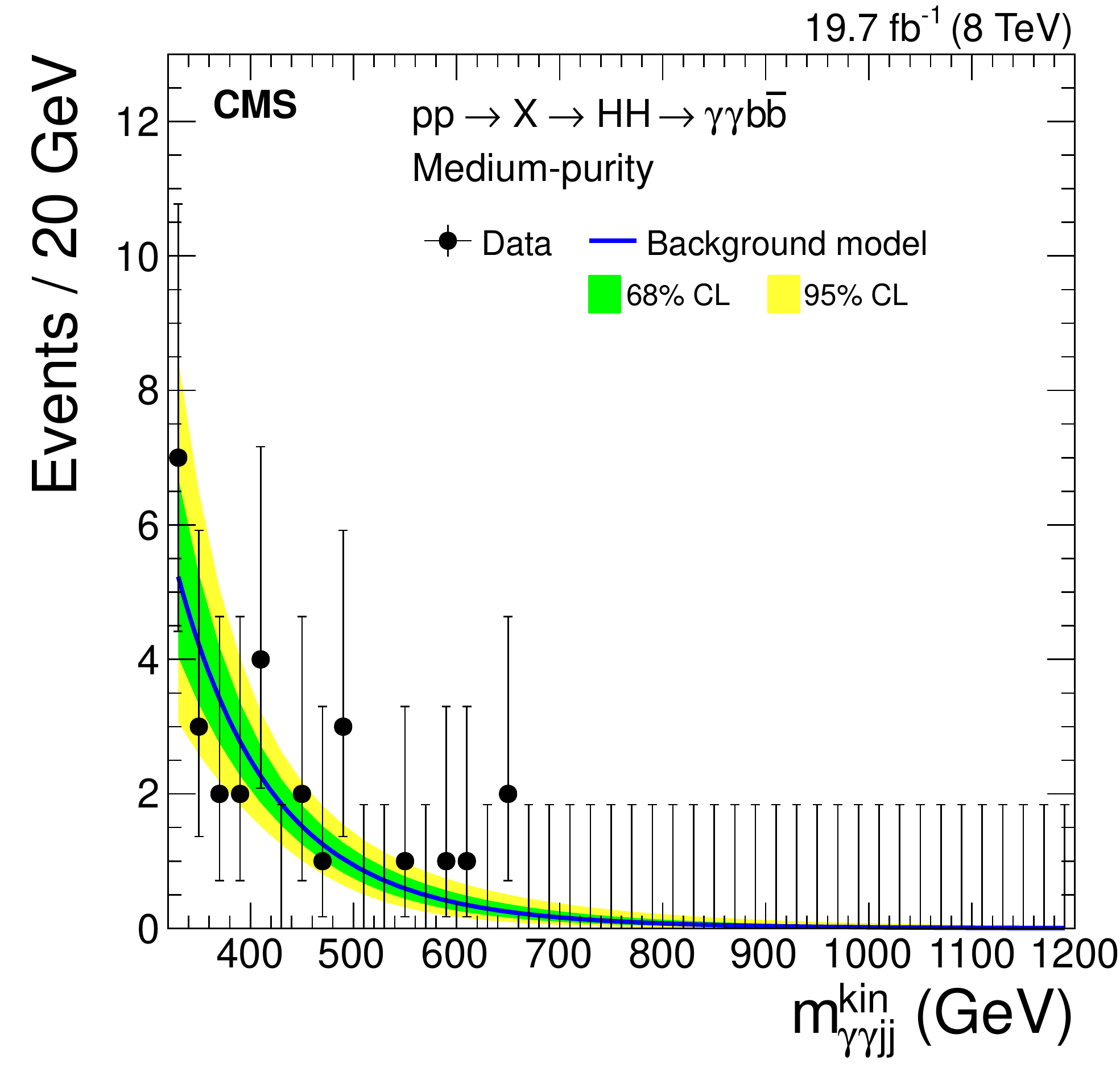}
\includegraphics[width=0.48\textwidth]{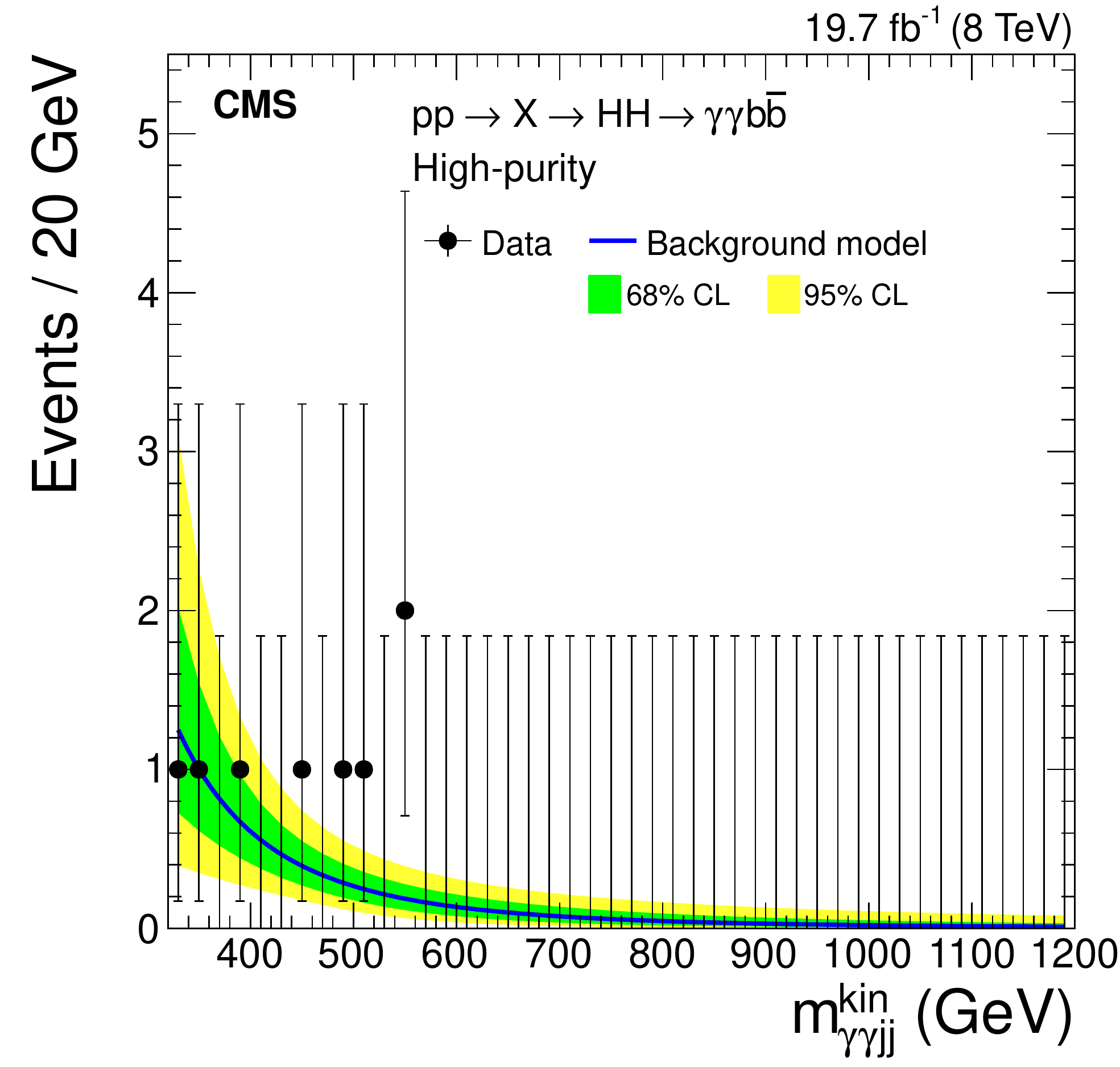}
\caption{High-mass resonant analysis: fits to the nonresonant background contribution to the $\Mggjjk$ spectrum in medium- (left) and in high-purity (right) events. The fits to the background-only hypothesis
are given by the blue curves, along with their 68\% and 95\% \CL contours.
}
\label{figure:BackgroundShapeMggjj}\end{figure*}

A possible signal can be extracted from a fit to the $\Mggjjk$ distribution for mass points between $320 \leq \Mggjjk \leq 1200\GeV$.
The background-only PD is a power law for each category, and is seen to well describe the data in Fig.~\ref{figure:BackgroundShapeMggjj}. The lower threshold  of 320\GeV is chosen to avoid the kinematic turn-on, while still ensuring full containment of signal for the $\mx \geq 400\GeV$ mass hypotheses.
Single-Higgs boson production is a negligible background in this
phase space region, and is absorbed into the parametrization of the nonresonant background.

\begin{table}[htb]
\renewcommand{\arraystretch}{1.1}
\topcaption{Additional selection criteria applied in the high-mass resonant search. \label{table:mass_cuts_high}}
\centering
\begin{scotch}{  ccc }
{Variable} & \multicolumn{2}{c}{Range (\GeVns)}\\ \cline{2-3}
& Medium-purity & High-purity \\ \hline
$\Mgg$  & 122--128 & 120--130 \\
$\Mjj$  & \multicolumn{2}{c}{85--170} \\
\end{scotch}
\end{table}

\subsection{Nonresonant}

We apply a selection on  $\acosthetastar$ in the search for nonresonant two-Higgs boson production.
To increase the sensitivity to a large variety of BSM topologies (see examples shown in Fig.~\ref{figure:massea_signal}), an additional categorization is applied in
$\Mggjjk$. For the SM-like topology in $\Pg\Pg \to \HH$ production, the $\Mggjjk$ spectrum peaks roughly at 400\GeV, while for
$\abs{\kapl} \gtrsim 10$ the peak shifts down
to the kinematic threshold of $\Mggjjk \approx 250\GeV$.
Large absolute values of the $\ctwo$ ($\abs{\ctwo} \approx 3$) parameter usually lead to an opposite effect by shifting the peak in $\Mggjjk$
spectrum above 400\GeV. Two categories are defined for $\Mggjjk$ smaller or larger than 350\GeV, a value optimized for SM-like search.
The details of the selections and categorizations are provided in Table~\ref{table:mass_cutsnonres}.

\begin{figure*}[htb]\centering
\includegraphics[width=0.48\textwidth]{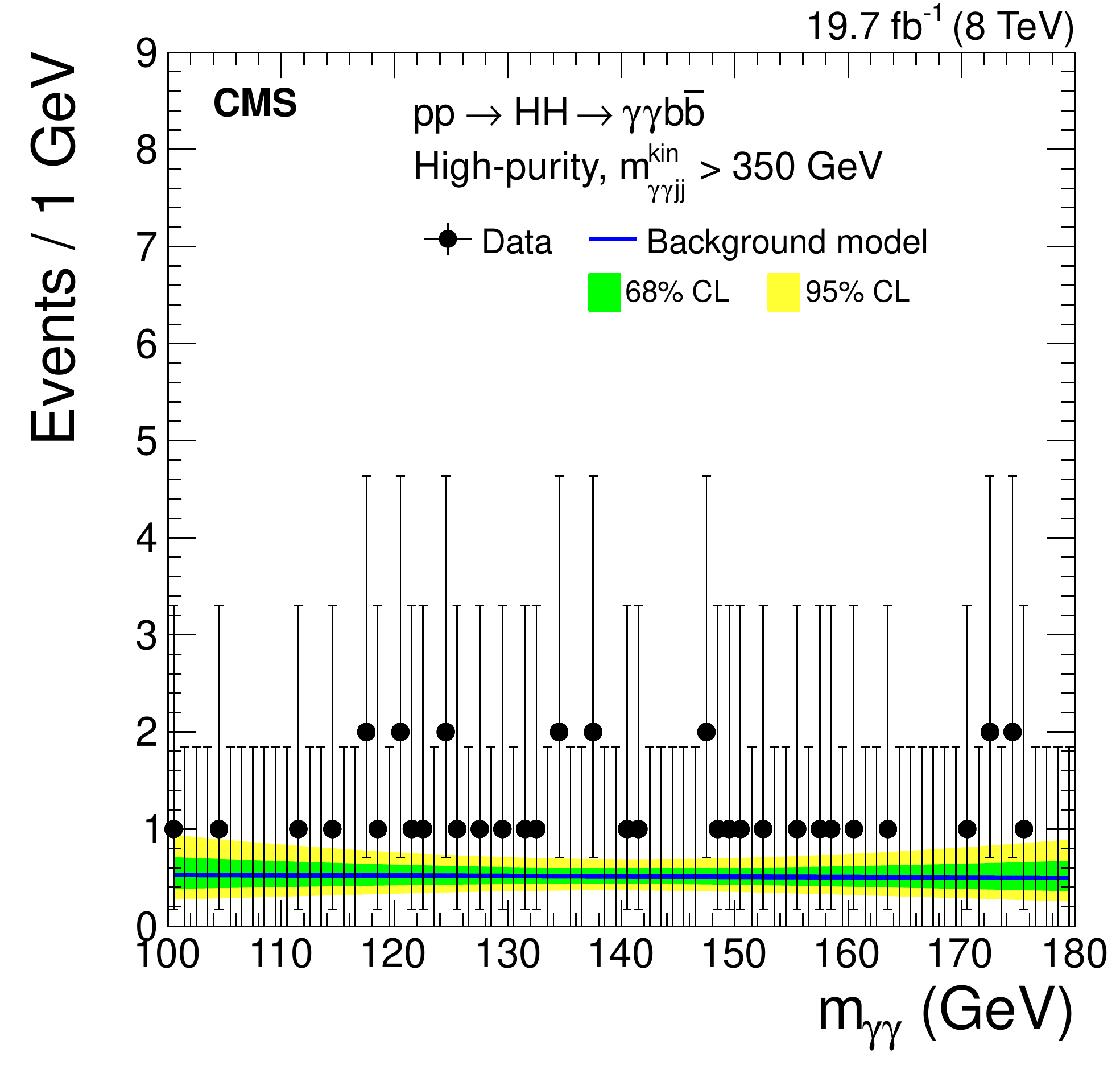}
\includegraphics[width=0.48\textwidth]{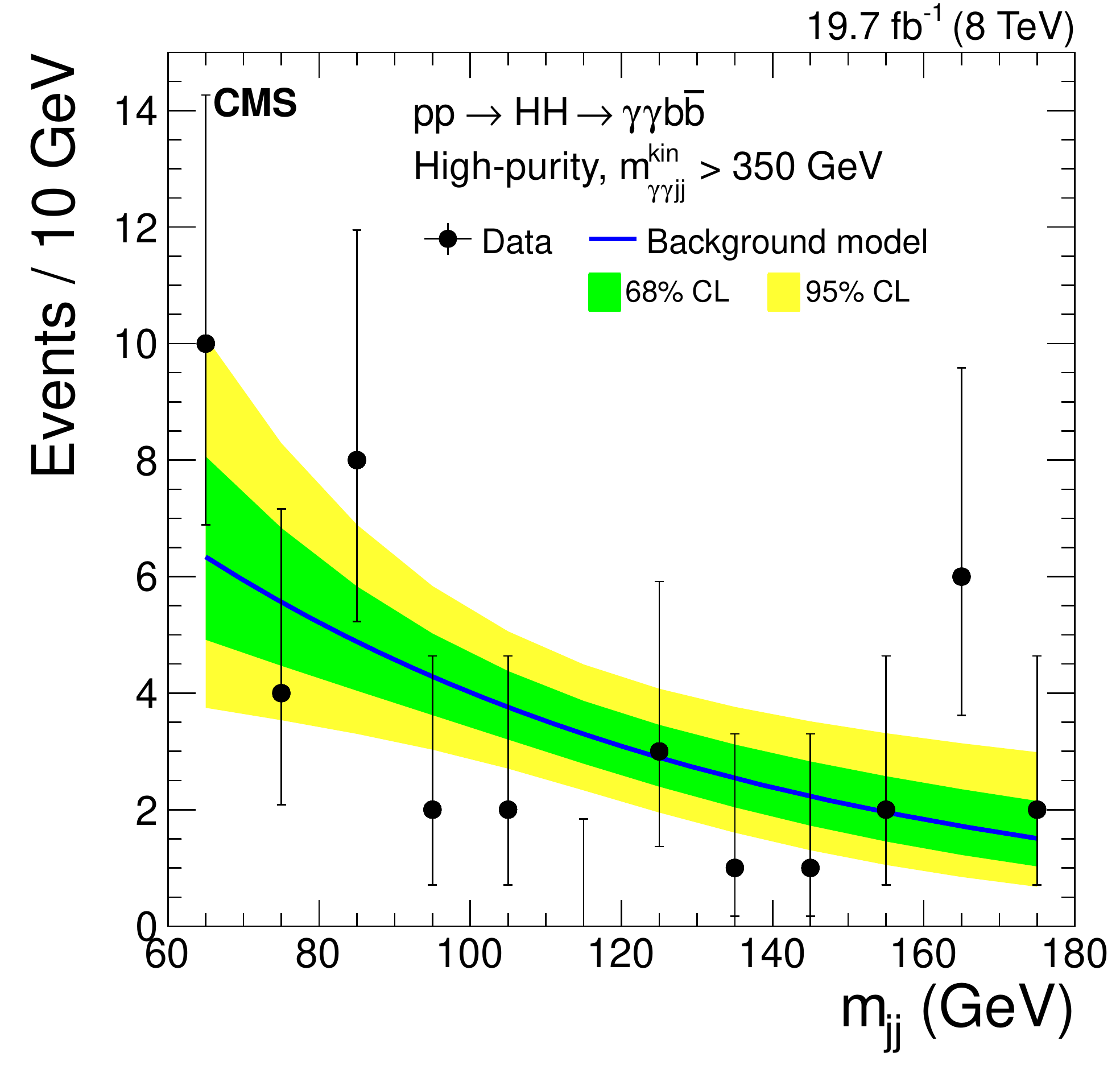}
\includegraphics[width=0.48\textwidth]{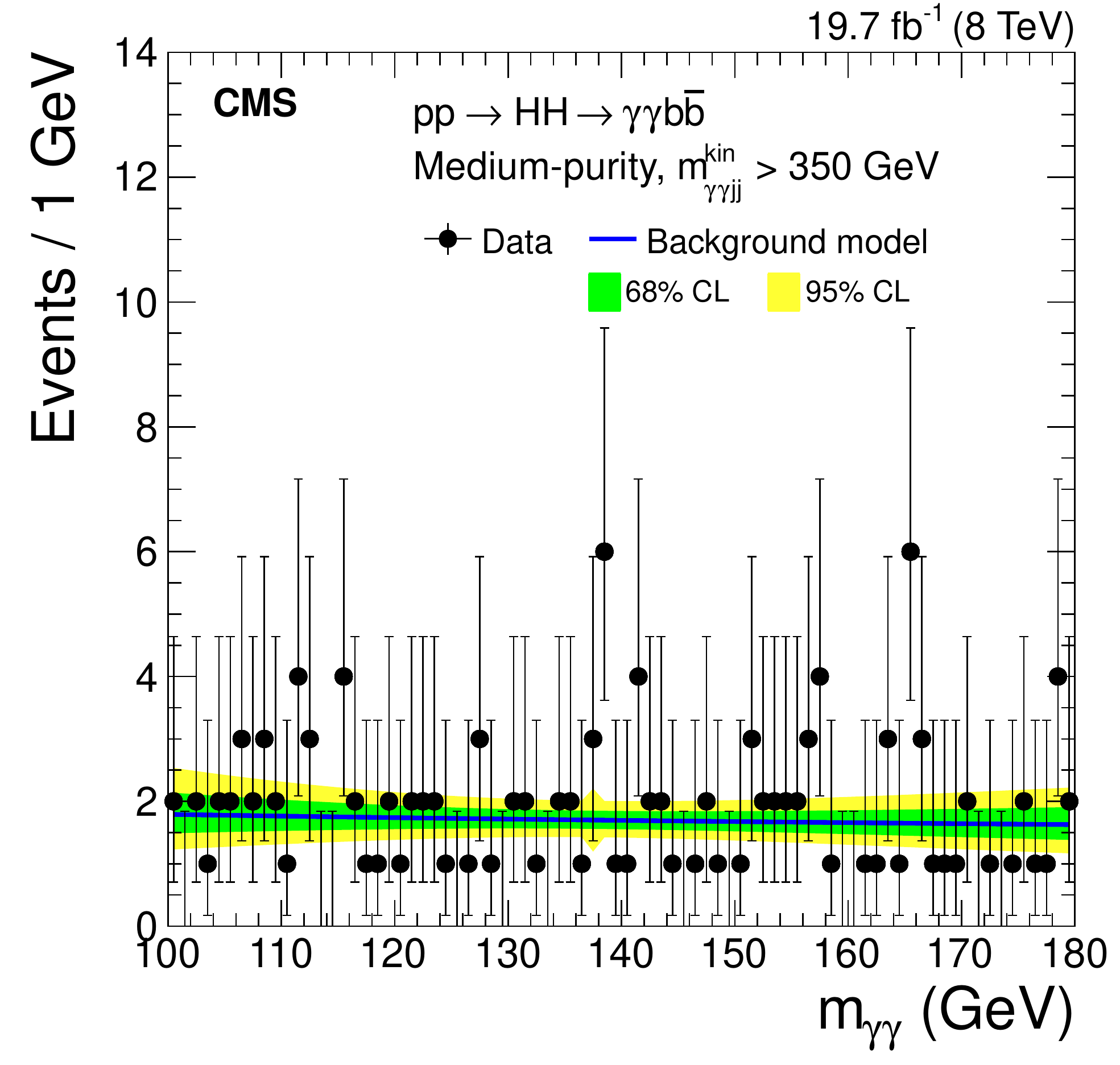}
\includegraphics[width=0.48\textwidth]{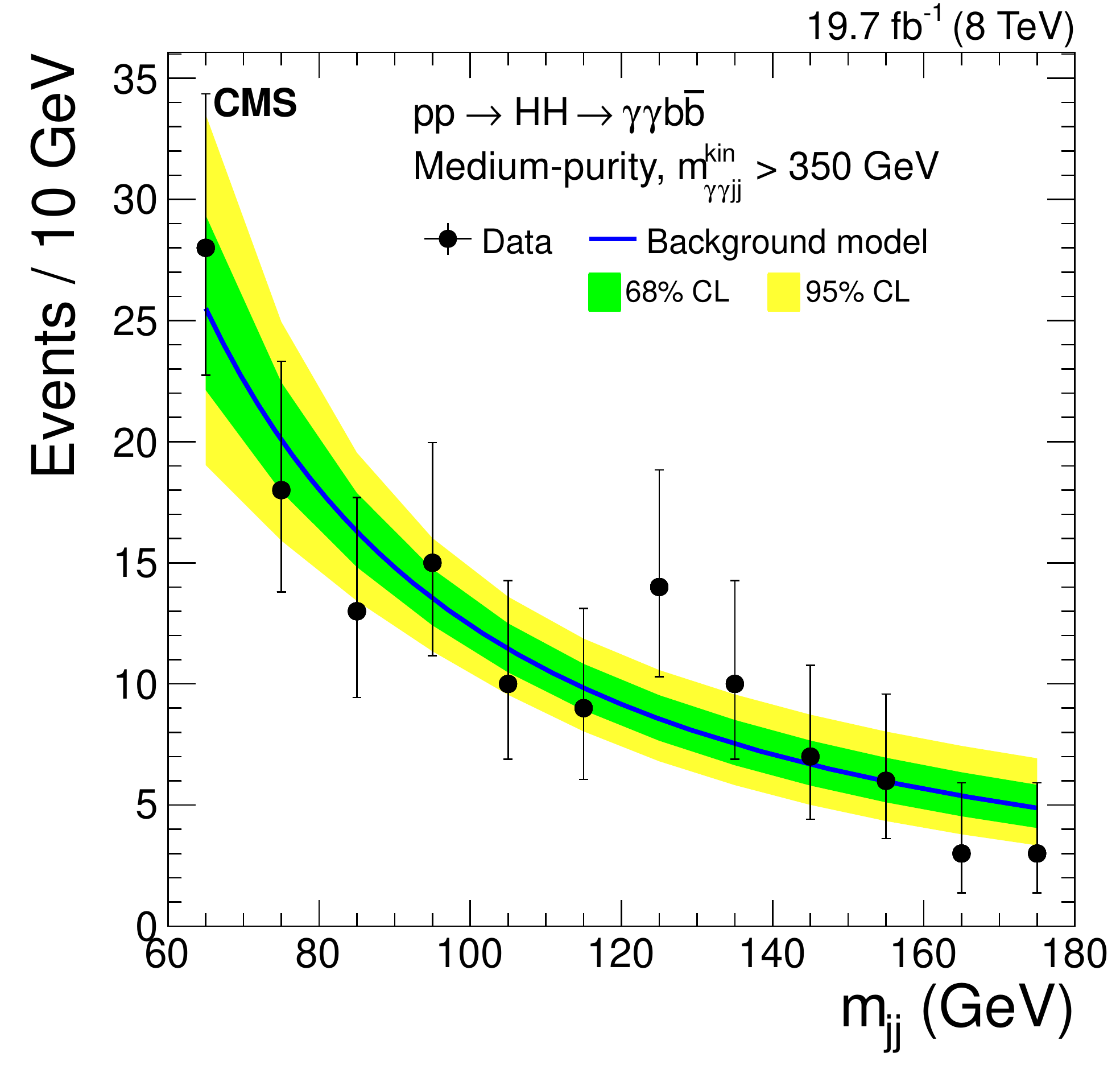}
\caption{Nonresonant analysis: fits to the nonresonant background contribution in high-$\Mggjjk$ and high-purity category to the $\Mgg$ (top-left) and $\Mjj$ spectra (top-right), and similarly for medium-purity category in bottom-left and bottom-right, respectively.
The fits to the background-only hypothesis
are given by the blue curves, along with their 68\% and 95\% \CL contours.
}
\label{figure:BackgroundShapeMggjjnon_mgg}\end{figure*}

\begin{figure*}[htb]\centering
\includegraphics[width=0.48\textwidth]{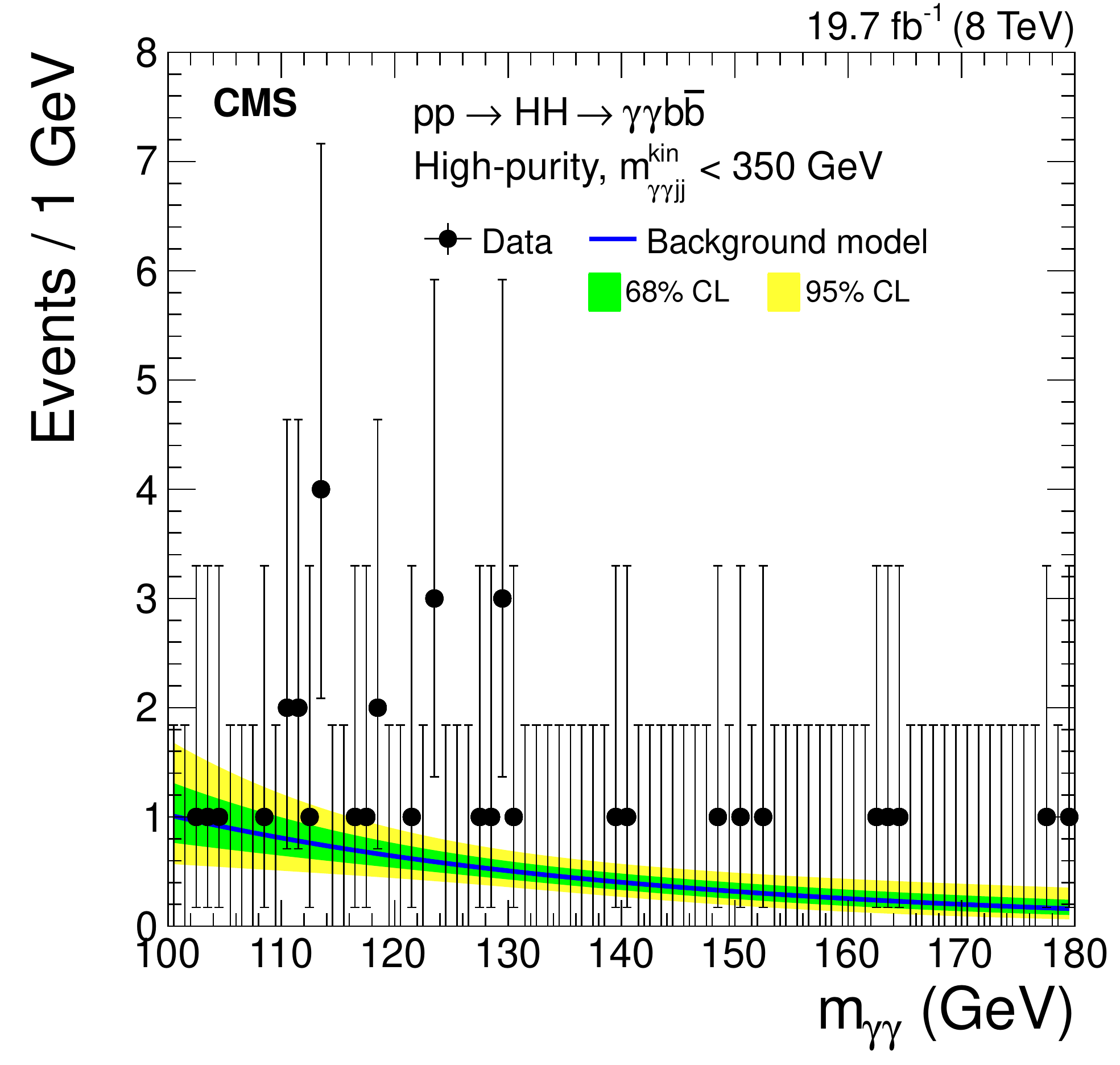}
\includegraphics[width=0.48\textwidth]{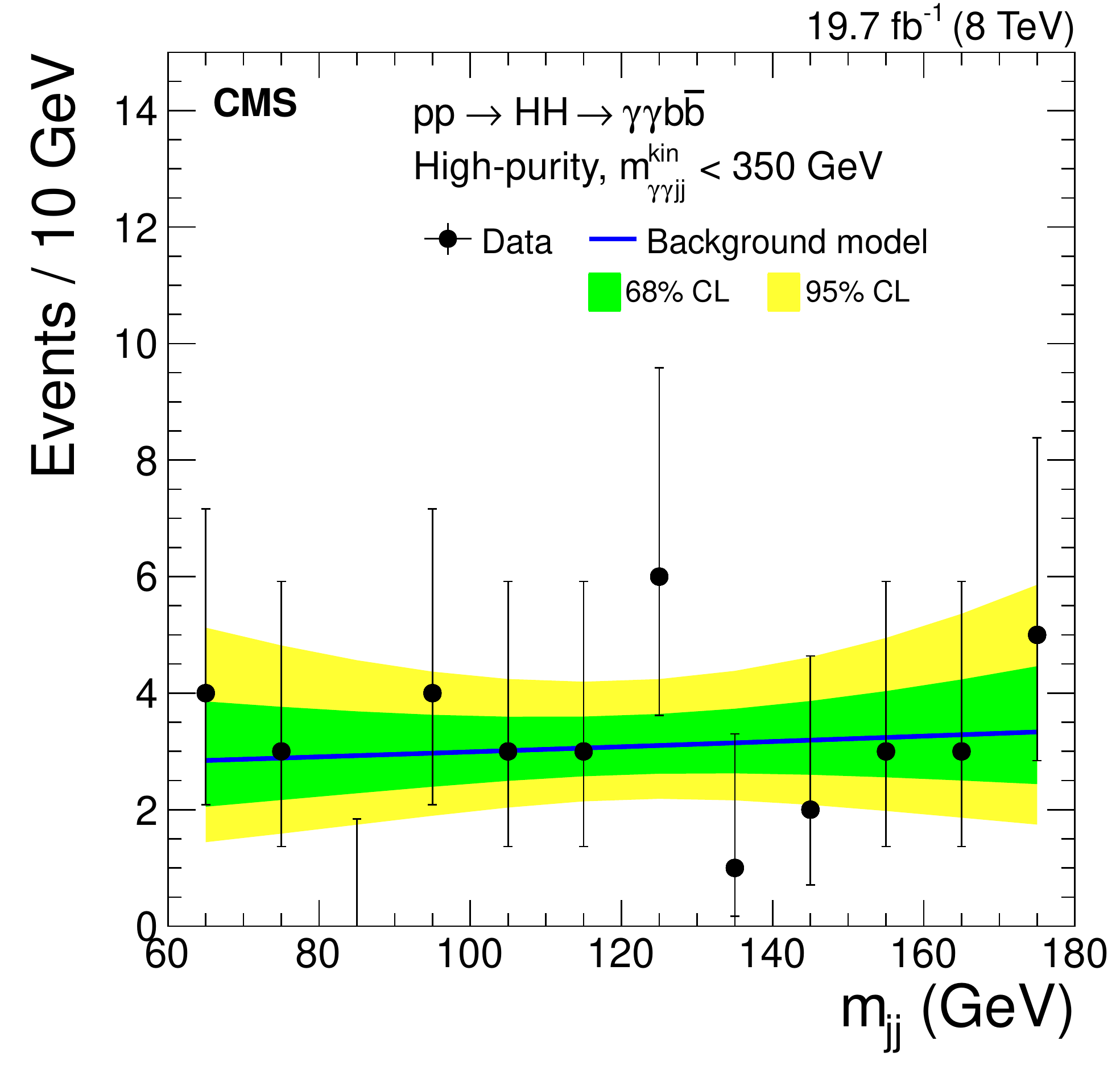}
\includegraphics[width=0.48\textwidth]{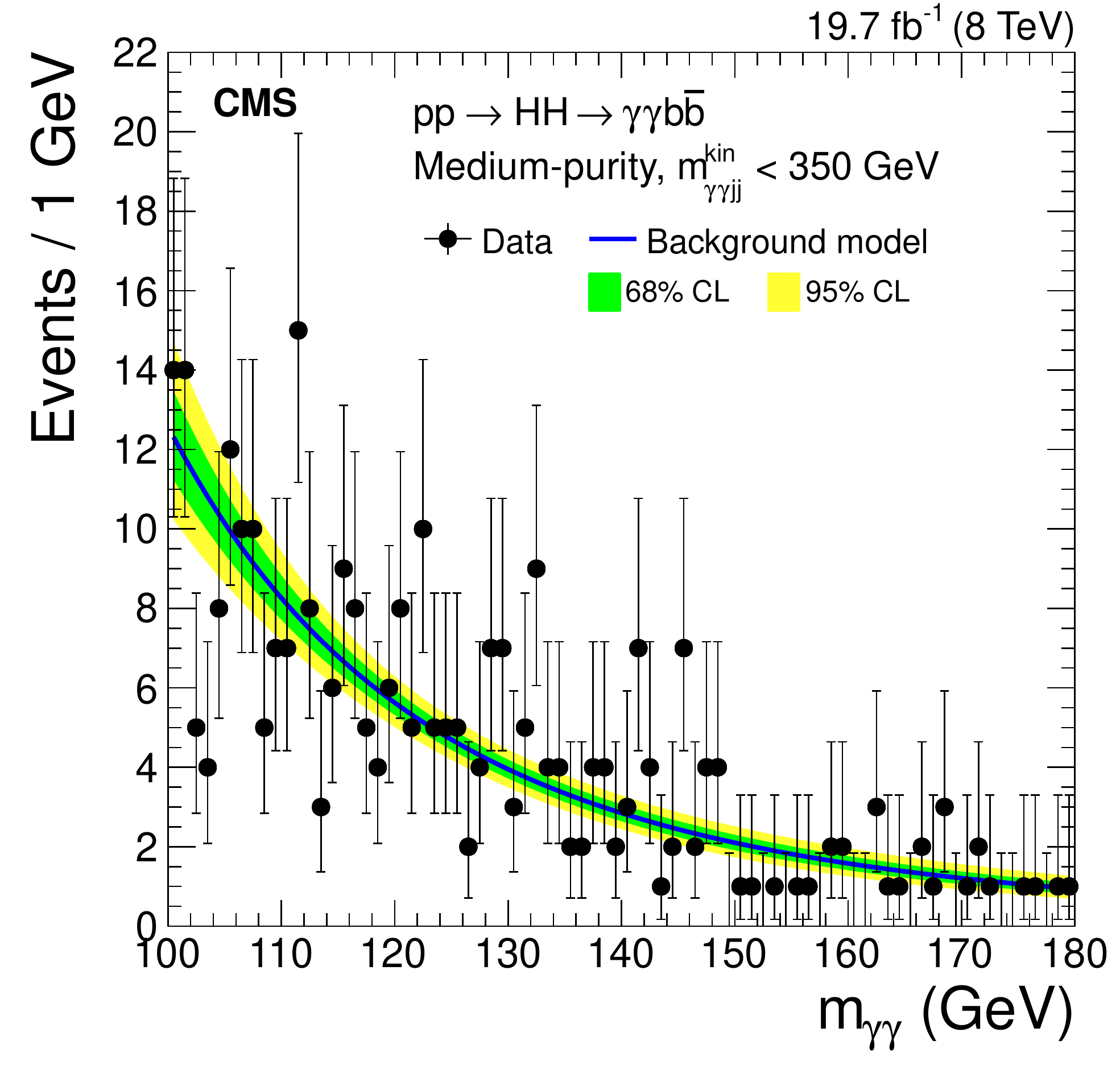}
\includegraphics[width=0.48\textwidth]{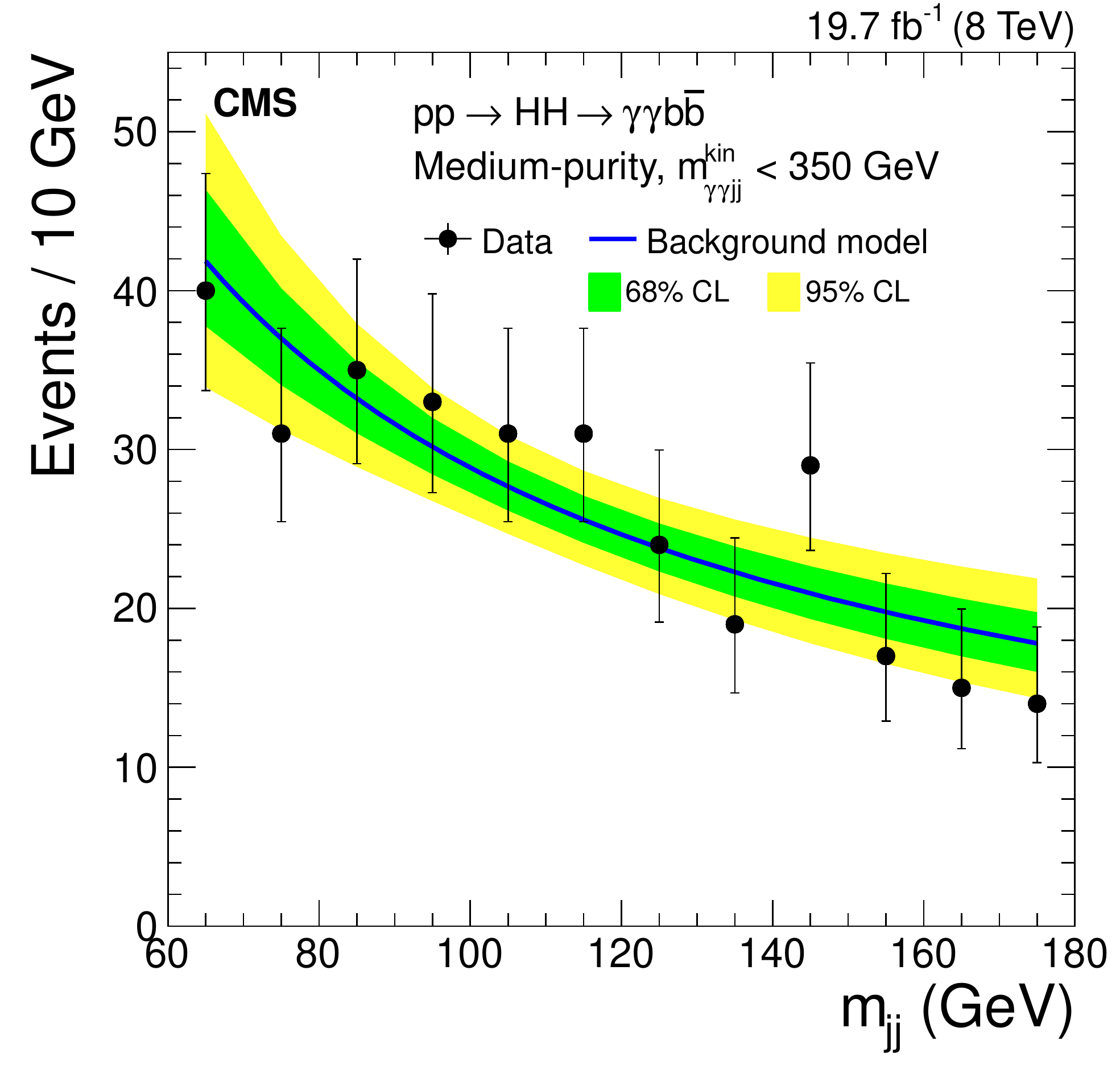}
\caption{
Nonresonant analysis: fits to the nonresonant background contribution in low-$\Mggjjk$ and high-purity category to the $\Mgg$ (top-left) and $\Mjj$ spectra (top-right), and similarly for medium-purity category in bottom-left and bottom-right, respectively.
The fits to the background-only hypothesis
are given by the blue curves, along with their 68\% and 95\% \CL contours.}
\label{figure:BackgroundShapeMggjjnon_mjj}\end{figure*}

A possible signal can be extracted using a simultaneous fit to the $\Mgg$ and
$\Mjj$ spectra. The background-only PD are exponentials and power-law expressions
for the medium- and high-purity categories, respectively, which agree with the data, as can be seen in Figs.~\ref{figure:BackgroundShapeMggjjnon_mgg} and~\ref{figure:BackgroundShapeMggjjnon_mjj}.

\begin{table*}[htb]
\renewcommand{\arraystretch}{1.3}
\topcaption{Additional selections applied in the nonresonant searches. \label{table:mass_cutsnonres}}
\centering
\begin{scotch}{ crlrl}
Variable & \multicolumn{2}{c}{High-purity} & \multicolumn{2}{c}{Medium-purity} \\\hline
$\acosthetastar$ &  \multicolumn{2}{c}{$<$0.9} & \multicolumn{2}{c}{$<$0.65} \\
$\Mggjjk$ categorization (\GeVns) & $<$350  & $>$350 & $<$350 & $>$350  \\
\end{scotch}
\end{table*}

\subsection{Signal efficiency}
\label{sec:signaleff}

\begin{figure}
\centering
\includegraphics[width=\cmsFigWidth]{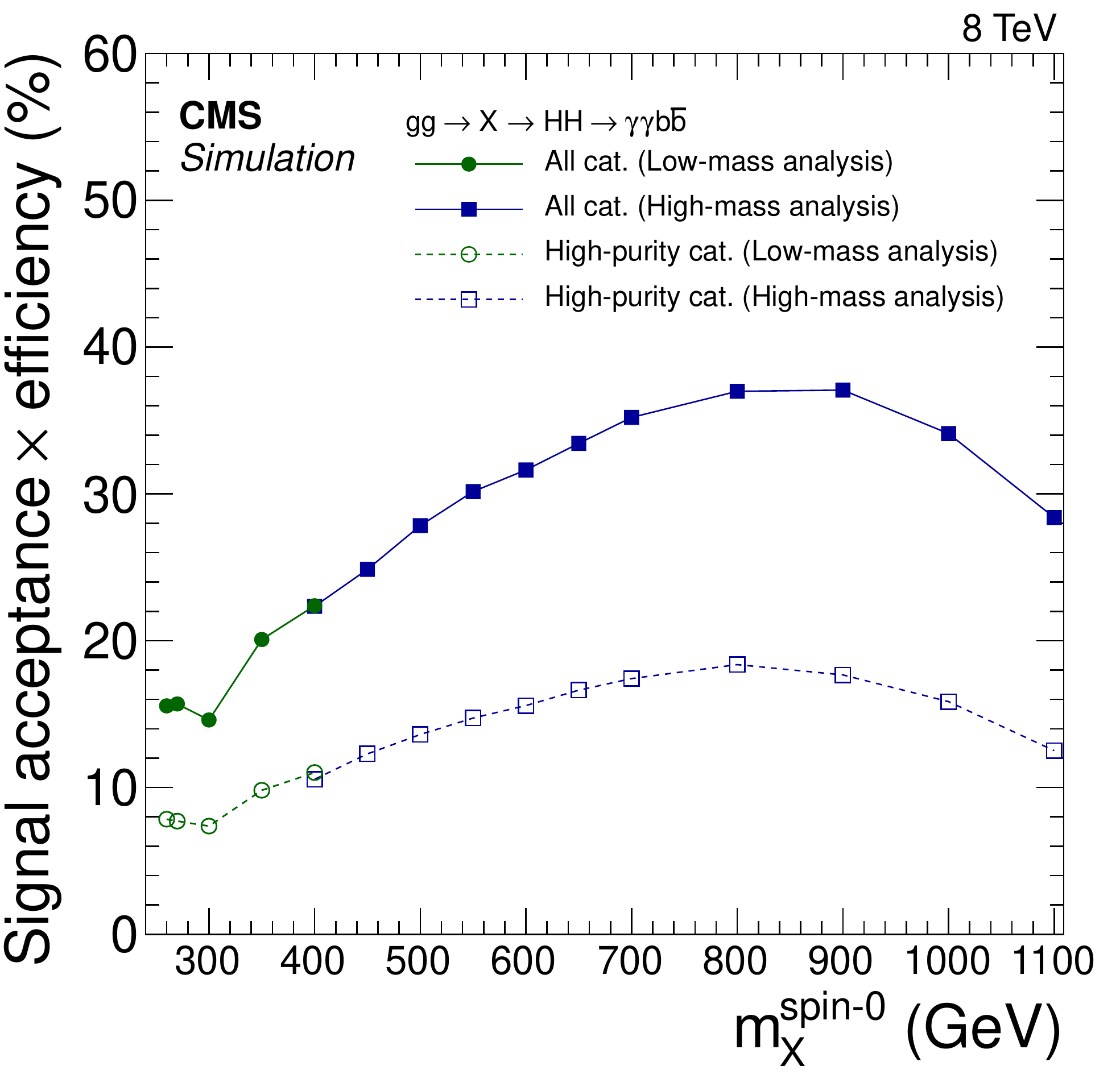}
\caption{Resonant signal efficiency for the final selection
  described in Table~\ref{table:gencut} and Section~\ref{sec::AnalysisMethods}. The efficiency is shown for a spin-0 hypothesis of a radion particle, but is similar for a spin-2 hypothesis of a KK graviton. The error bars associated with statistical uncertainties are smaller than the size of the markers.
\label{figure:efficiencies}}
\end{figure}

The signal efficiency is  a function of the mass hypothesis, as shown in
Fig.~\ref{figure:efficiencies}. It is estimated with respect to all events generated in a given signal sample.
The efficiency increases as
the resonance mass increases from $\mx =260$ to 900\GeV because of higher
photon and jet reconstruction efficiencies.
The efficiency starts to drop for $\mx > 900\GeV$. At this point,
the typical angular distance in the laboratory frame between two \PQb quarks produced in Higgs boson decay
is of the order of the distance parameter $D$~\cite{Gouzevitch:2013qca}. The minimum in efficiency is observed at $\mx = 300\GeV$.
It results from an optimization procedure designed to maximize the overall analysis sensitivity.
This procedure chooses an optimal size of $\Mggjjk$ window for each $\mx$ hypothesis.
For $\mx = 300\GeV$, the background is largest and the resulting $\Mggjjk$ window is smallest,
inducing a small drop in signal selection efficiency.
Finally, the single and double
\PQb tag categories contribute in roughly equal ways to the total efficiency.

\begin{figure}[htbp]\centering
\includegraphics[width=.49\textwidth]{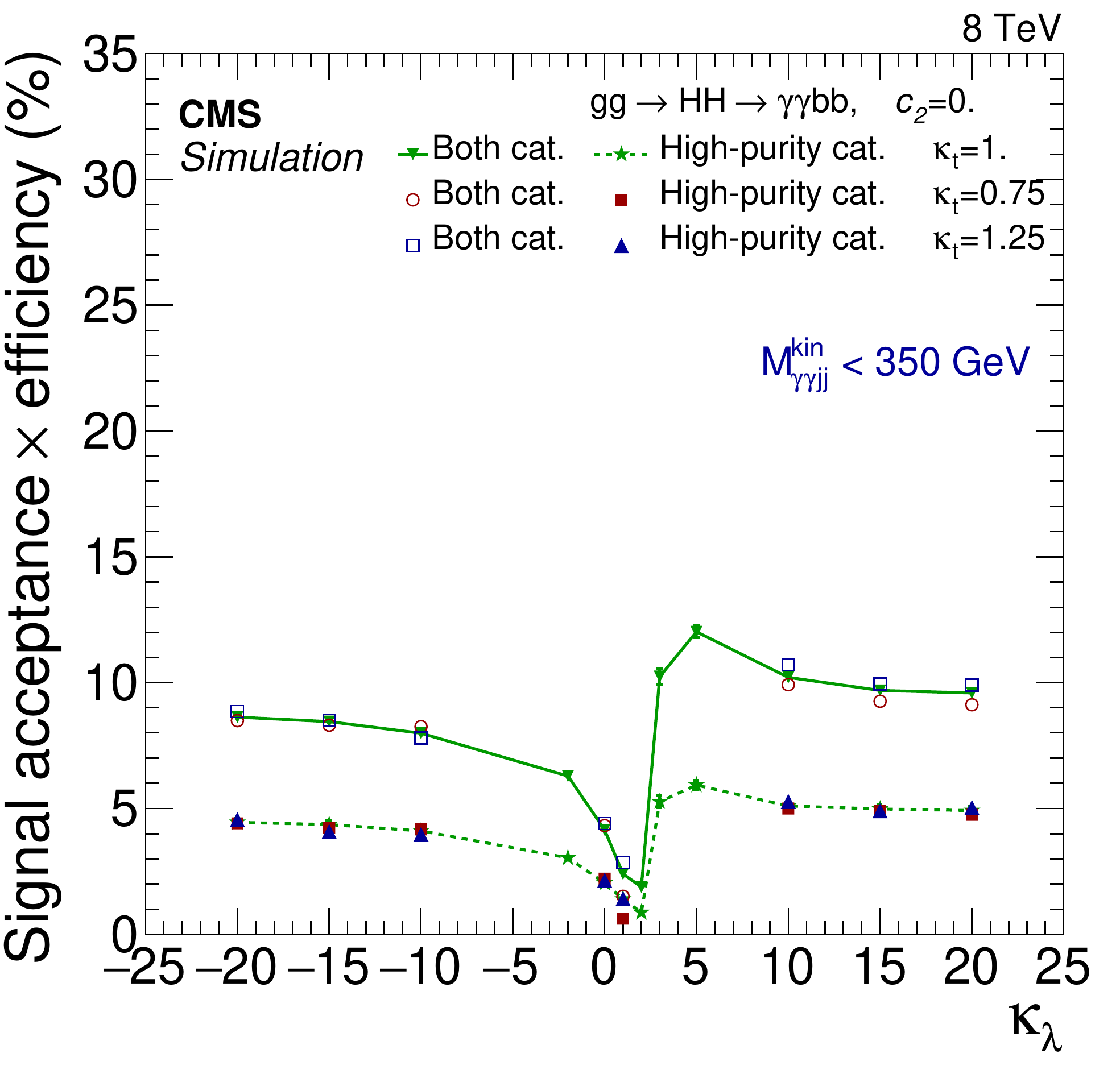}
\includegraphics[width=.49\textwidth]{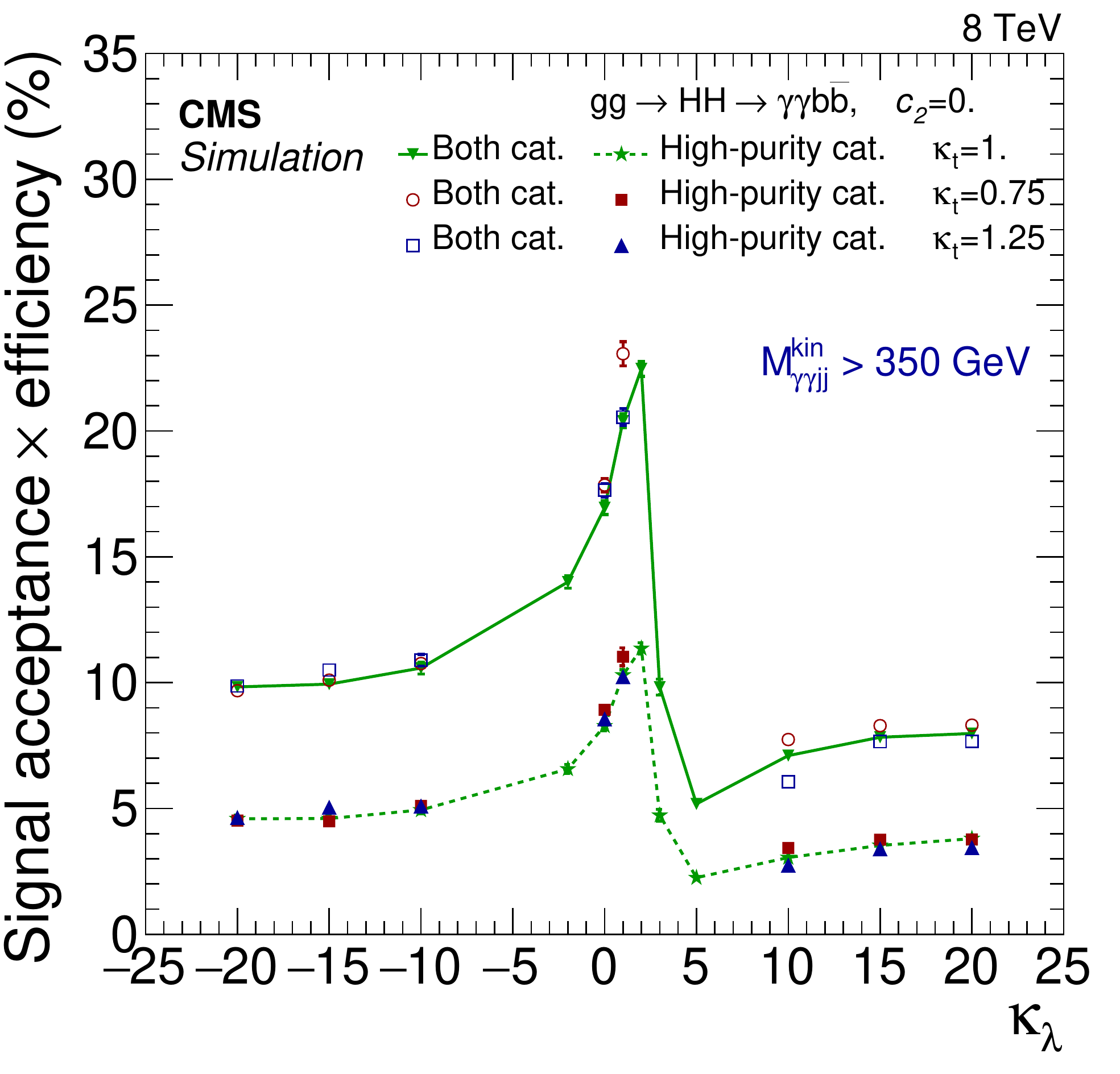}
\caption{
\label{fig:nonres_eff_4d_c2_0}
Signal efficiency for $\ctwo = 0$ as function of $\kapl$ for different values of $\kapt$, for the low-$\Mggjjk$ region (\cmsLeft) and high-$\Mggjjk$ region (\cmsRight).
}
\end{figure}

Figure \ref{fig:nonres_eff_4d_c2_0} provides the efficiencies of selecting the signal events as function of $\kapl$ for different values of $\kapt$ and assuming $\ctwo=0$. The \cmsLeft plot provides efficiencies for $\Mggjjk < 350\GeV$ categories and \cmsRight for $\Mggjjk > 350\GeV$ categories. For large values of $\abs{\kapl}$  (typically larger than 10) the efficiency is rather flat, while for small values of $\abs{\kapl}$ the efficiency in the $\Mggjjk < 350\GeV$ ($\Mggjjk > 350\GeV$) categories is reduced (increased). The change in efficiency is caused by the interference between two-Higgs box diagrams and the Higgs self coupling channel.
The total efficiency in four categories is $\approx$15--30\%, depending on the model parameters.
This figure illustrates the way that $\Mggjj$ categorization can help  separate different nonresonant signal hypotheses.

\clearpage
\section{Systematic uncertainties}
\label{section:sys}

The analysis defines a likelihood function based on the total PD and the data. The parameters
for total signal and for the background-only PD are constrained in the fit to maximize this function. A
uniform prior is used to parametrize the background PD. When converting the
fitted yields into production cross sections, we use simulations to estimate the selection efficiency for the signal. The difference between the simulation and the data is corrected through scaling factors. The uncertainty in those factors is taken into account through parameters included in the likelihood function. The nuisance parameters (parameters not of immediate interest)
are varied in the fit according to a log-normal probability density function. They can be classified into three categories.
The first category contains the uncertainty in the estimation of the integrated luminosity, which is taken as 2.6\%~\cite{CMS-PAS-LUM-13-001}. The second category includes systematic uncertainties
that modify the efficiency of signal selection. Finally the third category contains the uncertainties that impact the
signal or the Higgs boson PD.
More precisely, the values of the PD parameters are taken from fits to the MC simulation of signal and Higgs boson production. The systematic uncertainties affect the mean values and the resolution parameters of the PD, while all other CB parameters are fixed to their best values.
The sources of nuisance parameters are described below and their contribution to different categories are presented in Table \ref{table:espsyst}.

The photon-related uncertainties are discussed in Ref.~\cite{Khachatryan:2014ira}.
While the photon energy scale (PES) is known at
the sub-percent level in the region of $\pTg$ characteristic
of the SM $\PH \to\gamma\gamma$ signal, the uncertainty increases to 1\% for $\pTg>100\GeV$.
The photon energy resolution (PER) is known with a 5\% precision~\cite{Khachatryan:2014ira}.
A 1\% normalization uncertainty is estimated in the offline diphoton selection efficiency
and in the trigger efficiency.  An additional  normalization
uncertainty of 5\% is estimated for the high-mass region to account
for differences in the \pt spectrum of signal photons and of electrons
from $\mathrm{Z} \to \Pe\Pe$ production used to estimate the quoted uncertainties.

The uncertainty in the jet energy scale (JES) is accounted for by changing the jet response by 1--2\% \cite{CMS-DP-2013-011},
depending on the kinematics, while the uncertainty in the jet energy resolution (JER)
is estimated by changing the jet resolution by
10\% \cite{JINST6}.  An additional 1\% uncertainty in the four-body mass
accounts for effects in the high-mass region related to the partial overlap
between the two \PQb jets from the Higgs boson decay.
The uncertainty in the \PQb tagging efficiency is estimated by changing
the b tagging scale factor up and down by one standard deviation in each purity category~\cite{BTV}.
The related systematic uncertainties are known to be anticorrelated between the two categories.

Theoretical systematic uncertainties are considered for the single-Higgs boson contribution from SM production,
corresponding to the scale dependence of higher-order terms and impact from the choice of proton
parton distribution functions (PDF)~\cite{Dittmaier:2011ti, Heinemeyer:2013tqa}.
No theoretical uncertainties are assumed on BSM
signals.
However, there is one exception.
We consider the situation
where the kinematic properties of the new signal are
identical to those of the SM, but the cross section is different (SM-like search).
In that case we parametrize the BSM cross section $\sigmaHH^\mathrm{BSM}$ by the ratio $\mu_{\rm \HH} = \sigmaHH^\mathrm{BSM}/\sigmaHH^\mathrm{SM}$. When such a search is performed the theoretical uncertainties on $\sigmaHH^\mathrm{SM}$ are included in the likelihood.
Finally, an additional systematic uncertainty of 0.24\GeV is assigned to account for the experimental uncertainty in the Higgs boson mass~\cite{Aad:2015zhl}.
The impact of this uncertainty is comparable to the one from PES.

The analysis is limited by the statistical precision. The systematic uncertainties worsen the expected
cross section limits by at most 1.5 and 3.8\% in the resonant and nonresonant searches, respectively.

\begin{table*}
\renewcommand{\arraystretch}{1.3}
\topcaption{
Summaries of systematic uncertainties. For the normalization uncertainties, the values in the right column indicate the impact on the signal normalization. The uncertainty in the \PQb tagging efficiency is anticorrelated between the \PQb tag categories. The uncertainty in the $\Mggjjk$ categorization is anticorrelated between $\Mggjjk$ categories for the nonresonant search. \label{table:espsyst}}
\centering
\begin{scotch}{lr}
\multicolumn{2}{c}{General uncertainties in normalization} \\
\hline
Integrated luminosity & 2.6\%\\
Diphoton trigger efficiency & 1.0\% \\
Diphoton selection efficiency & 1.0\% \\
\hline
\multicolumn{2}{c}{ {Resonant low-mass and nonresonant analyses: 2D fit to $\Mgg$ and $\Mjj$}} \\
\hline
\multicolumn{2}{c}{---------------- Uncertainties in normalization ----------------} \\[\cmsTabSkip]
Acceptance in $\pTj$ ( JES and JER) & 1.0\%\\
\PQb tagging efficiency in the high-purity category & 5.0\% \\
\multicolumn{2}{l}{\PQb tagging efficiency in the medium-purity category} \\
\tab Low-mass resonant and nonresonant $\Mggjjk < 350\GeV$ & 2.1\% \\
\tab Nonresonant $\Mggjjk > 350\GeV$ & 2.8\% \\
\multicolumn{2}{l}{$\Mggjjk$ acceptance (PES, JES, PER and JER)}\\
\tab Low-mass resonant & 1.5\%\\
\tab Nonresonant $\Mggjjk < 350\GeV$ categories & 1.5\%\\
\tab Nonresonant $\Mggjjk > 350\GeV$ categories & 0.5\%\\[\cmsTabSkip]
\multicolumn{2}{c}{---------------- Uncertainties in the PD parameters ----------------} \\[\cmsTabSkip]
$\Mjj$ resolution (JER), $\frac{\Delta \sjjG}{\sjjG}$ and $\frac{\Delta \sjjCB}{\sjjCB}$& 10\% \\
$\Mjj$ scale (JES),      $\frac{\Delta \Mujj}{\Mujj}$ & 2.6\%\\
$\Mgg$ resolution (PER), $\frac{\Delta \sggG}{\sggG}$ and $\frac{\Delta \sggCB}{\sggCB}$& 5\% \\
\multicolumn{2}{l}{$\Mgg$ scale (PES and uncertainty in $\mH$)}       \\
\tab Low-mass resonant, $\frac{\Delta \Mugg}{\Mugg}$ &  0.4\%\\
\tab Nonresonant, $\frac{\Delta \Mugg}{\Mugg}$ & 0.5\%\\
\hline
\multicolumn{2}{c}{High-mass resonant analysis: 1D fit to $\Mggjjk$} \\
\hline
\multicolumn{2}{c}{---------------- Uncertainties in normalization ----------------} \\[\cmsTabSkip]
\PQb tagging efficiency in the high-purity category & 5.0\% \\
\PQb tagging efficiency in the medium-purity category & 2.8\% \\
$\Mjj$ and $\pTj$ acceptance related to JES and JER & 1.5\%\\
$\Mgg$ selection acceptance related to PES and PER & 0.5\% \\
Extra high $\pTg$ normalization uncertainty & 5.0\% \\[\cmsTabSkip]
\multicolumn{2}{c}{---------------- Uncertainties in the PD parameters ----------------} \\[\cmsTabSkip]
$\Mggjjk$ scale (PES and JES), $\frac{\Delta \Muggjjk}{\Muggjjk}$ & 1.4\% \\
$\Mggjjk$ resolution (PER and JER), $\frac{\Delta \sggjjkG}{\sggjjkG}$ and $\frac{\Delta \sggjjkCB}{\sggjjkCB}$ & 10.0\% \\
\end{scotch}
\end{table*}

\section{Results}
\label{section:results}

No significant excess is observed over the background expectation in the resonant or nonresonant searches.
Upper limits are computed using the modified frequentist approach for confidence levels (CL$_s$), taking the profile likelihood
as a test statistic~\cite{CLS1,CLS2} in the asymptotic approximation. The limits are subsequently compared
to theoretical predictions assuming SM branching fractions for Higgs boson decays.

\subsection{Resonant signal}

\begin{figure}[htbp]\centering
\includegraphics[width=0.48\textwidth]{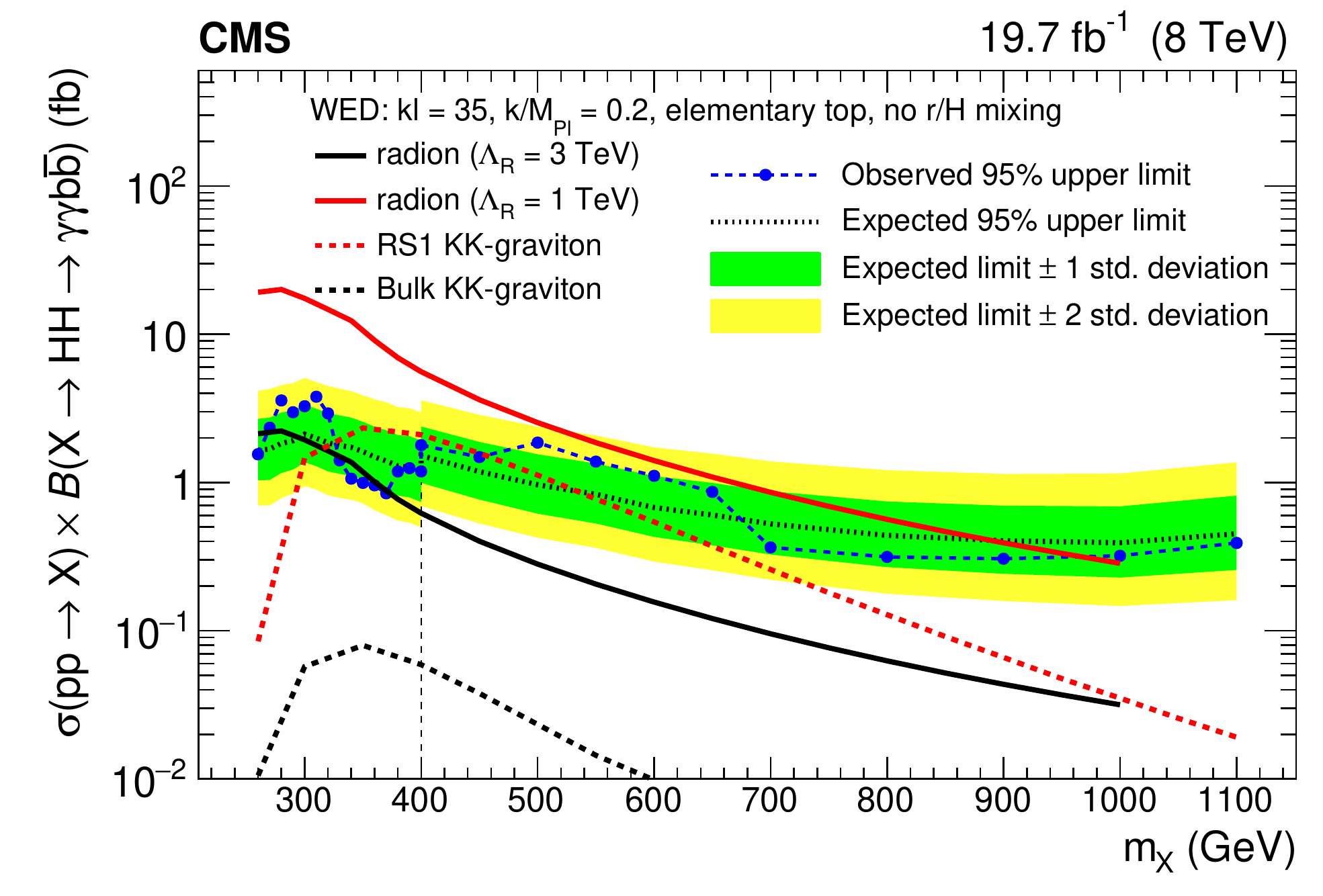}
\includegraphics[width=0.48\textwidth]{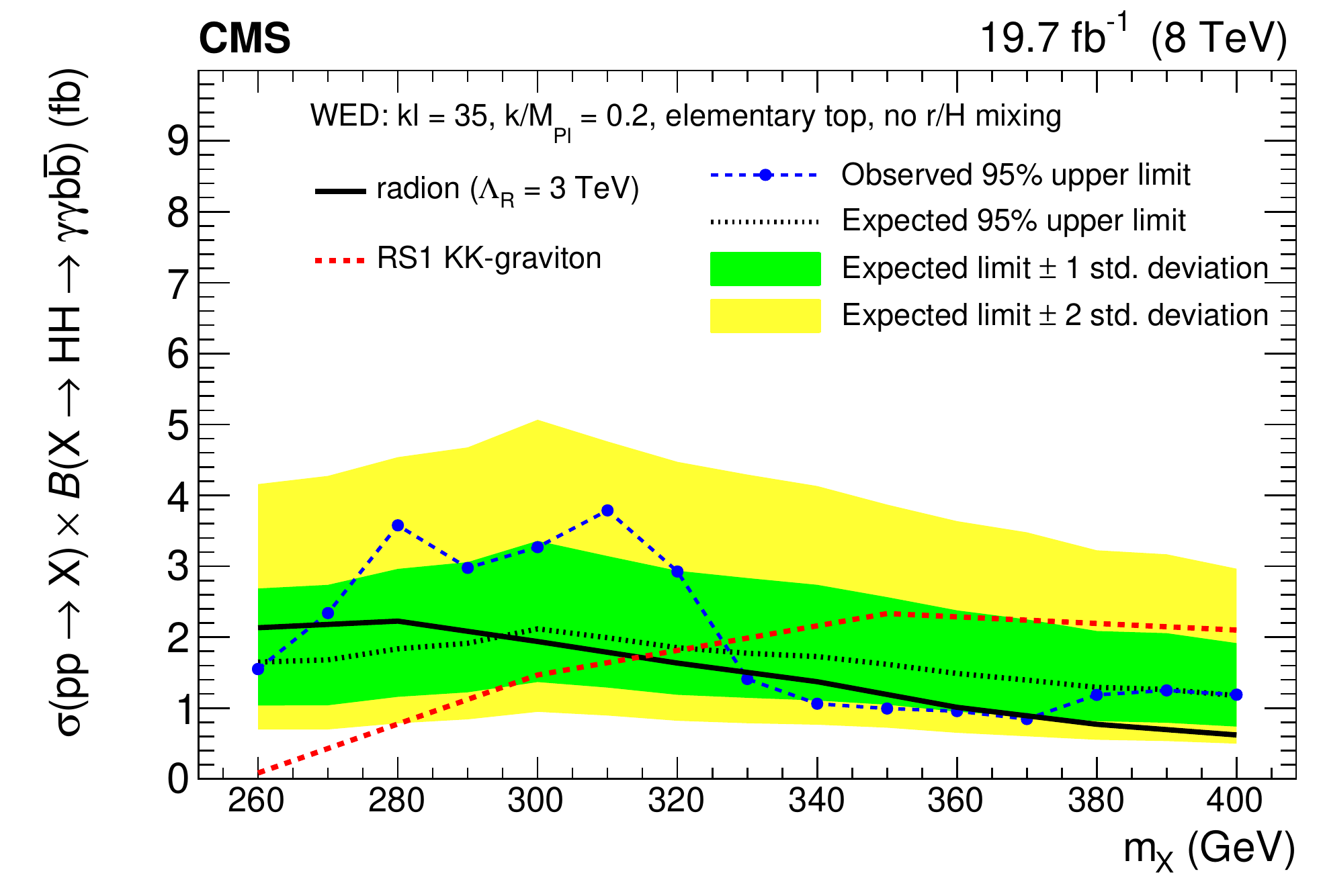}
\caption{
Observed and expected 95\% \CL upper limits on the product of cross section and the
  branching fraction $\sigma(\Pp\Pp \to \mathrm{X}) \, \mathcal{B}(\mathrm{X} \to \HH \to
  \gamma\gamma \bbbar)$ obtained through a combination of the two event categories
  (\cmsLeft), and in the zoomed view at low-mass (\cmsRight).  The green and yellow bands represent, respectively, the
1 and 2 standard deviation extensions beyond the expected
limit.  Also shown are theoretical predictions corresponding to WED models for radions and RS1 KK gravitons. The upper plot with a logarithmic scale for the y-axis also
provides the prediction for the production cross section of a bulk KK graviton.
The vertical dashed line in the upper plot shows the separation between the low-mass and high-mass analyses. The limits for $\mx = 400\GeV$ are shown for both methods.
}
\label{figure:ExpectedLimits}\end{figure}

The observed and median expected upper limits for all the data at 95\% \CL
are shown in the top of Fig.~\ref{figure:ExpectedLimits}, and at the bottom in a zoomed-in view of the
low-mass region.
The expected limits range from 1.99\unit{fb} for $\mx = 310\GeV$ to 0.39\unit{fb} for $\mx = 1\TeV$. At the transition
point between the low-mass and high-mass searches, $\mx = 400\GeV$, results with both methods are provided.
An improvement of about 20\% is observed from the use of the 2D model approach with respect to the 1D analysis.

The result is compared with the cross sections for KK-graviton and radion production in WED models.
The tools used to calculate the cross sections for the production of KK graviton in the bulk and RS1 models are described in Refs.~\cite{Agashe:2013kyb, deAquino:2011ix}. The implementation of the calculations is described in Ref.~\cite{Oliveira:2014kla}.
In analogy with the Higgs boson, the radion field is predominantly produced through gluon-gluon fusion~\cite{Mahanta:2000zp, Davoudiasl:2000wi}.
The cross section for radion production is calculated  at NLO electroweak and next-to-next-to-leading logarithmic QCD
accuracy, using the recipe suggested in Ref.~\cite{Giudice:2000av}. This recipe consists of multiplying the radion cross section based on the
fundamental parameter of the theory, $\LambdaR$, by a $K$-factor calculated for SM-like Higgs boson
production through gluon-gluon fusion~\cite{Catani:2003zt,Heinemeyer:2013tqa}. The calculations are performed for the SM-like Higgs boson with masses up to 1\TeV. We use the CTEQ6L PDF~\cite{Nadolsky:2008zw} in these calculations. No mixing between a radion and the Higgs boson is considered in this paper.

In Table~\ref{table:res}, we summarize the inclusive production cross sections and the branching fractions of the
heavy resonances in the theoretical benchmarks we use for interpretation.
The absolute values of the production cross sections scale with $(k/\AMpl)^2$ for the KK Graviton~\cite{Oliveira:2010uv} and with $1/\LambdaR^2$ for the radion~\cite{Barger:2011qn}.

The values for the branching fractions of the resonances in the theory benchmarks do not depend on the
fundamental parameters of the theory.
The resonance decays have a phase space suppression,
related to the mass difference between the resonance and its decay products.
In this way, the decay to a Higgs boson pair is not allowed if $\mx<250\GeV$ nor
to top quark pairs if $\mx<350\GeV$.
In Table~\ref{table:res}, we see that the value of the branching fraction changes with the resonance mass from $\mx=300$ to $\mx=500\GeV$.
The exact pattern of this phenomenon is related to the balance between the different phase space suppressions for decays to $\HH$ or to \ttbar, which depends on the model under consideration.

\begin{table*}[htpb]
\caption{Cross section and branching fractions for the benchmark theories used in this paper~\cite{Oliveira:2010uv,Barger:2011qn}. The branching fractions does not depend on $k/\AMpl$, nor on $\LambdaR$.}
\label{table:res}
\centering
\begin{scotch}{cccc}

Model & $\mx$ (\GeVns) & $\sigma (\Pg\Pg \to \mathrm{X})$ (pb) & $\mathcal{B}(\mathrm{X}\to \HH)$ \\[1mm]\hline
RS1 KK graviton & 300 & 2140 & 0.03\% \\
($k/\AMpl =0.2$) & 500 & 172  & 0.24\% \\
 & 1000 & 3.1 & 0.43\% \\[\cmsTabSkip]
Bulk-RS KK graviton & 300 & 0.65  & 0.89\% \\
($k/\AMpl =0.2$) & 500 & 0.11  & 8.2\% \\
 & 1000 & 0.0021  & 9.8\% \\[\cmsTabSkip]
Radion & 300 & 20.7  & 32\% \\
($\LambdaR = 1$\TeV) & 500 & 3.87 & 25\%\\
 & 1000 & 0.46 & $24$\% \\
\end{scotch}
\end{table*}

The analysis excludes a radion with masses below 980\GeV for the radion scale
$\LambdaR=1\TeV$.
The search has also sensitivity to the presence of a radion with an ultraviolet cutoff $\LambdaR =3\TeV$ in the region between $200$ and $300\GeV$.

The difference in total selection efficiency between the spin-0
(radion) and the spin-2 (KK-graviton) models does not exceed 3\%.
Thus, the same upper limits that are extracted using a radion simulation
can be used directly to exclude a KK graviton with
masses between 325 and 450\GeV, assuming $k/\AMpl = 0.2$.
The analysis is not yet sensitive to the presence of a  KK graviton in the bulk scenario with the same parameters.

\subsection{Nonresonant signal}

We consider the kinematic properties for new signal
identical to those of the SM, but with a different cross section.
The observed and expected upper limits on SM-like
$\Pg\Pg \to \HH \to \gamma \gamma \bbbar$ production are, respectively, 1.85 and 1.56\unit{fb}. This can be translated into
0.71 and 0.60\unit{pb}, respectively, for the total  $\Pg\Pg \to \HH$ production cross section.
The results can also be interpreted in terms of observed and expected limits on the scaling factor $\mu_{\HH} < 74$
and ${<}62_{-22}^{+37}$, respectively. This result provides a quantification  of the current analysis relative to the SM prediction.

We also interpret the results in the context of Higgs boson anomalous couplings.
The cross section for nonresonant two-Higgs-boson production $\sigmaHH^\mathrm{BSM}$ in this context can
be written as a polynomial in the parameters of the theory relative to the
SM nonresonant cross section $\sigmaHH^\mathrm{SM}$ as:
\ifthenelse{\boolean{cms@external}}{
\begin{multline}
\frac{\sigmaHH}{\sigmaHH^\mathrm{SM}} =
A_1\, \kapt ^4 + A_2\, \ctwo^2 + A_3\, \kapt^2\, \kapl^2\\
 + ( A_6\, \ctwo + A_7\, \kapt \kapl )\kapt^2
+ A_8\, \kapt \kapl \ctwo.
\label{eq:nonrescx}
\end{multline}
}{
\begin{equation}
\frac{\sigmaHH}{\sigmaHH^\mathrm{SM}} =
A_1\, \kapt ^4 + A_2\, \ctwo^2 + A_3\, \kapt^2\, \kapl^2
 + ( A_6\, \ctwo + A_7\, \kapt \kapl )\kapt^2
+ A_8\, \kapt \kapl \ctwo.
\label{eq:nonrescx}
\end{equation}
}
The numerical  coefficients of Eq.~(\ref{eq:nonrescx}) can be calculated by fitting cross sections
as described in Ref.~\cite{Dall'Osso:2015aia}, obtaining thereby:
$A_1 = 2.19,$ $A_2= 9.9,$ $A_3= 0.324,$ $A_6 = -8.7,$ $A_7 = -1.51,$  and $A_8 = 3.0$.
Under the assumption that radiative corrections to gluon-gluon fusion of two-Higgs-bosons
do not depend significantly on anomalous interactions~\cite{Grober:2015cwa, deFlorian:2015moa}, we
normalize $\sigmaHH$
such that, when
$\kapt = 1$, $\kapl = 1$, and $\ctwo = 0$, to the cross section that equals the SM prediction
at NNLO in QCD.

\begin{figure*}[htb]\centering
\includegraphics[width=0.85\textwidth]{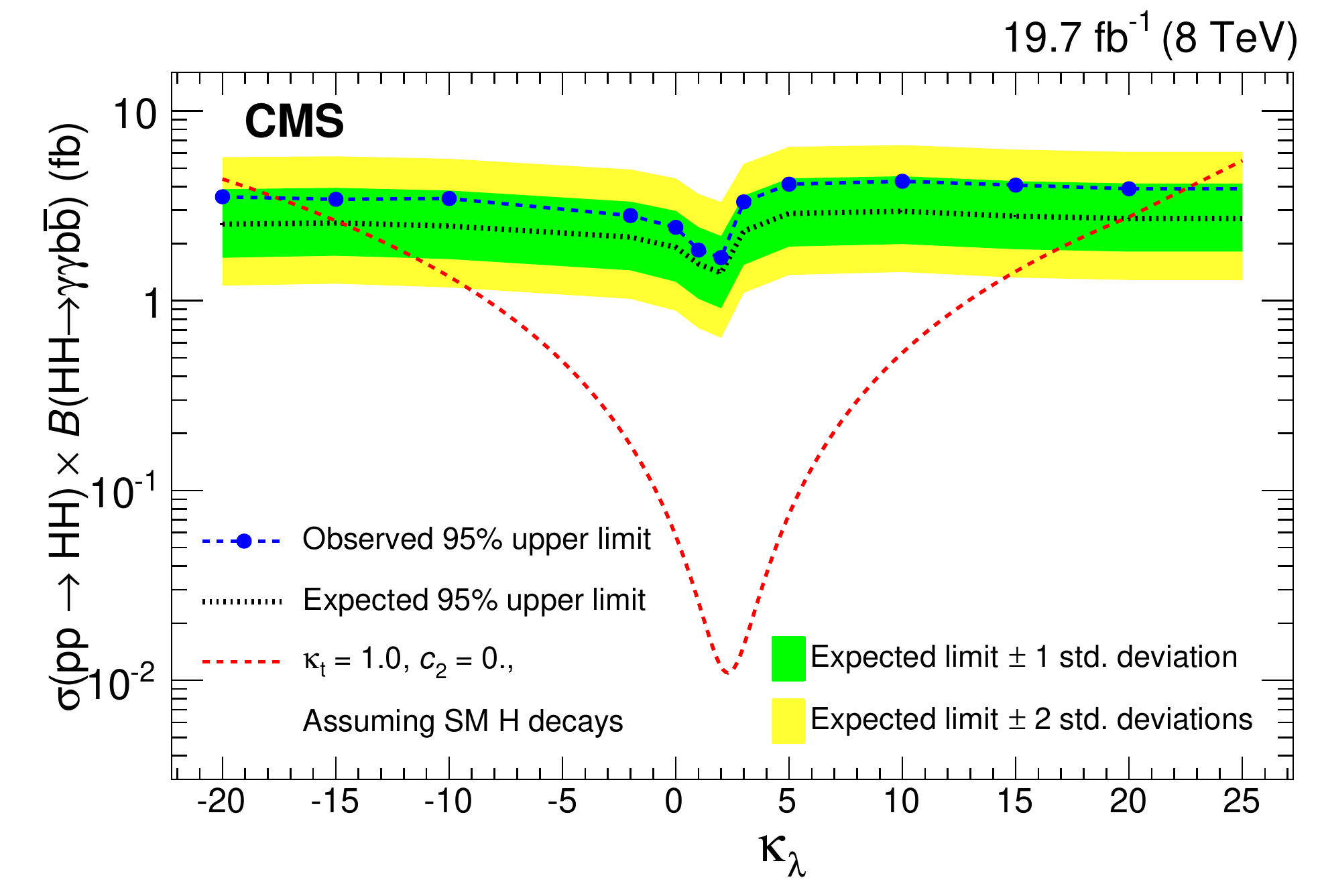}
\caption{Observed and expected 95\% \CL upper limits on the product of cross section and the
  branching fraction $\sigma (\Pp\Pp \to \HH) \, \mathcal{B}(\HH \to \gamma\gamma \bbbar)$
  for the nonresonant BSM analysis, performed by changing only
  $\kapl$, while keeping all other parameters fixed at the SM predictions.
  }
\label{figure:ExpectedLimitsNonRes}\end{figure*}

In Fig.~\ref{figure:ExpectedLimitsNonRes},
95\% \CL limits on nonresonant cross sections are shown, assuming changes
only in the trilinear Higgs boson couplings, with
the other parameters fixed to their SM values.
All $\kapl$ values are excluded below
$-17.5$ and above 22.5. These results are obtained by extrapolating the
limits between the simulated points, as well as above the highest
simulated value of $\kapl$ using Eq. \ref{eq:nonrescx}, which relies on the similarity
of distributions for signal at large values of $\abs{\kapl}$~\cite{Dall'Osso:2015aia, Carvalho:2016rys},
as well as on the behavior of the signal efficiency described in Section~\ref{sec:signaleff}.

\begin{figure*}[htb]\centering
\includegraphics[width=0.95\textwidth]{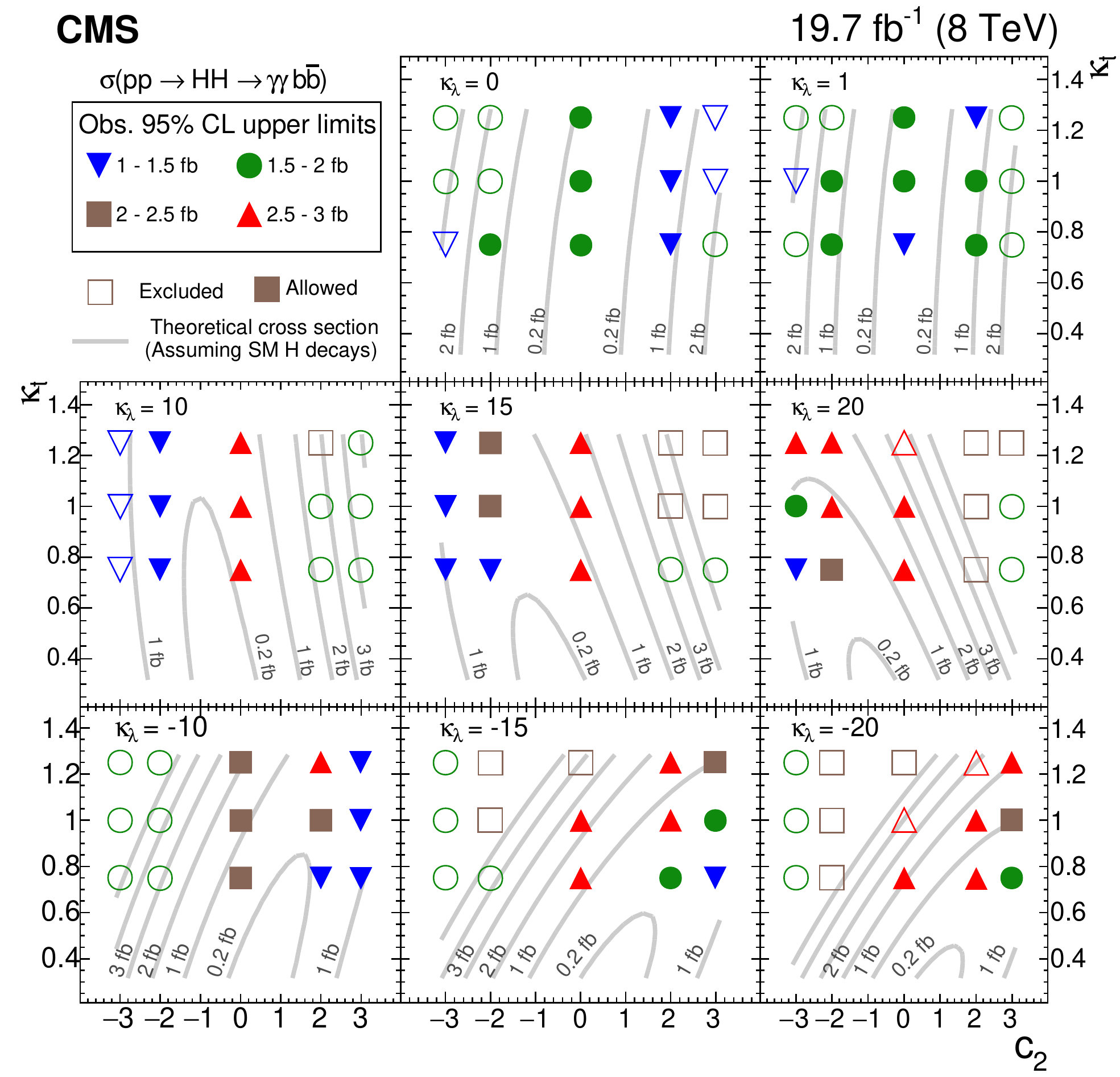}
\caption{
The observed 95\% \CL limits for nonresonant two-Higgs production in the $\ctwo$
and $\kapt$ planes for different values of $\kapl$.
The different markers symbolize the range in which the upper limits
in the cross sections are relevant.
The results are
compared to the theoretical prediction.
The gray lines
represent contours of equal cross section, as calculated using Eq.~(\ref{eq:nonrescx}).
The boxed-in cross section markers provide the combination of parameters excluded at 95\% \CL.
}
\label{figure:ExpectedLimitsLowMass}
\end{figure*}

Figure~\ref{figure:ExpectedLimitsLowMass} shows the 95\% \CL limits for nonresonant two-Higgs production in the $\ctwo$
and $\kapt$ planes for different values of $\kapl$.
The specific interference pattern for each combination of parameters
produces different exclusion limits for different simulated points of parameter space~\cite{Dall'Osso:2015aia, Carvalho:2016rys}.
Only discrete values are provided for limits because a linear interpolation between
the simulated points could not follow the strong variations due to interference terms.
The points in the theoretical phase space excluded by the data are surrounded by small black
boxes. Certain combinations of $\ctwo$, $\kapl$, or $\kapt$ parameters can be excluded under the assumption
that Higgs bosons have their usual SM branching fractions. For example, we observe that $\abs{\ctwo} \geq 3$ is
disfavored by the data when $\kapl$ and $\kapt$ are fixed to SM values.

\section{Summary}

A search is performed by the CMS collaboration for resonant and nonresonant
production of two Higgs bosons in the decay channel $\HH \to \gamma\gamma \bbbar$, based on an integrated
luminosity of 19.7\fbinv of proton-proton collisions collected at $\sqrt{s}=8\TeV$.
The observations are compatible with expectations from standard model processes.
No excess is observed over background predictions.

Resonances are sought in the mass range between 260 and 1100\GeV.
Upper limits at a 95\% \CL are extracted on cross
sections for the production of new particles decaying to Higgs boson pairs.
The limits are compared to BSM predictions, based
on the assumption of the existence of a warped extra dimension.
A radion with an ultraviolet cutoff $\LambdaR =1\TeV$ is excluded with
masses below 980\GeV.
The search has sensitivity to the presence of a radion with an ultraviolet cutoff $\LambdaR =3\TeV$ when its mass lies between 200 and 300\GeV.
The RS1 KK graviton is excluded with masses between 325 and 450\GeV for $k/\AMpl = 0.2$.
The analysis is not yet sensitive to the presence of a KK graviton in the bulk scenario with the same parameters.

For nonresonant production with SM-like kinematics, a 95\% \CL upper limit of 1.85\unit{fb} is set for the product of the HH cross section and branching fraction,
corresponding to a factor 74 larger than the SM value.
When only the trilinear Higgs boson coupling  is changed,
values of the self coupling are excluded for
$\kapl < -17$ and $\kapl > 22.5$.
The parameter space is also probed for the presence of other anomalous Higgs boson couplings.

\begin{acknowledgments}
We are grateful to B. Hespel, F. Maltoni, E. Vryonidou, and M. Zaro for a customized model of the nonresonant signal generation.

\hyphenation{Bundes-ministerium Forschungs-gemeinschaft Forschungs-zentren} We congratulate our colleagues in the CERN accelerator departments for the excellent performance of the LHC and thank the technical and administrative staffs at CERN and at other CMS institutes for their contributions to the success of the CMS effort. In addition, we gratefully acknowledge the computing centres and personnel of the Worldwide LHC Computing Grid for delivering so effectively the computing infrastructure essential to our analyses. Finally, we acknowledge the enduring support for the construction and operation of the LHC and the CMS detector provided by the following funding agencies: the Austrian Federal Ministry of Science, Research and Economy and the Austrian Science Fund; the Belgian Fonds de la Recherche Scientifique, and Fonds voor Wetenschappelijk Onderzoek; the Brazilian Funding Agencies (CNPq, CAPES, FAPERJ, and FAPESP); the Bulgarian Ministry of Education and Science; CERN; the Chinese Academy of Sciences, Ministry of Science and Technology, and National Natural Science Foundation of China; the Colombian Funding Agency (COLCIENCIAS); the Croatian Ministry of Science, Education and Sport, and the Croatian Science Foundation; the Research Promotion Foundation, Cyprus; the Ministry of Education and Research, Estonian Research Council via IUT23-4 and IUT23-6 and European Regional Development Fund, Estonia; the Academy of Finland, Finnish Ministry of Education and Culture, and Helsinki Institute of Physics; the Institut National de Physique Nucl\'eaire et de Physique des Particules~/~CNRS, and Commissariat \`a l'\'Energie Atomique et aux \'Energies Alternatives~/~CEA, France; the Bundesministerium f\"ur Bildung und Forschung, Deutsche Forschungsgemeinschaft, and Helmholtz-Gemeinschaft Deutscher Forschungszentren, Germany; the General Secretariat for Research and Technology, Greece; the National Scientific Research Foundation, and National Innovation Office, Hungary; the Department of Atomic Energy and the Department of Science and Technology, India; the Institute for Studies in Theoretical Physics and Mathematics, Iran; the Science Foundation, Ireland; the Istituto Nazionale di Fisica Nucleare, Italy; the Ministry of Science, ICT and Future Planning, and National Research Foundation (NRF), Republic of Korea; the Lithuanian Academy of Sciences; the Ministry of Education, and University of Malaya (Malaysia); the Mexican Funding Agencies (CINVESTAV, CONACYT, SEP, and UASLP-FAI); the Ministry of Business, Innovation and Employment, New Zealand; the Pakistan Atomic Energy Commission; the Ministry of Science and Higher Education and the National Science Centre, Poland; the Funda\c{c}\~ao para a Ci\^encia e a Tecnologia, Portugal; JINR, Dubna; the Ministry of Education and Science of the Russian Federation, the Federal Agency of Atomic Energy of the Russian Federation, Russian Academy of Sciences, and the Russian Foundation for Basic Research; the Ministry of Education, Science and Technological Development of Serbia; the Secretar\'{\i}a de Estado de Investigaci\'on, Desarrollo e Innovaci\'on and Programa Consolider-Ingenio 2010, Spain; the Swiss Funding Agencies (ETH Board, ETH Zurich, PSI, SNF, UniZH, Canton Zurich, and SER); the Ministry of Science and Technology, Taipei; the Thailand Center of Excellence in Physics, the Institute for the Promotion of Teaching Science and Technology of Thailand, Special Task Force for Activating Research and the National Science and Technology Development Agency of Thailand; the Scientific and Technical Research Council of Turkey, and Turkish Atomic Energy Authority; the National Academy of Sciences of Ukraine, and State Fund for Fundamental Researches, Ukraine; the Science and Technology Facilities Council, UK; the US Department of Energy, and the US National Science Foundation.

Individuals have received support from the Marie-Curie programme and the European Research Council and EPLANET (European Union); the Leventis Foundation; the A. P. Sloan Foundation; the Alexander von Humboldt Foundation; the Belgian Federal Science Policy Office; the Fonds pour la Formation \`a la Recherche dans l'Industrie et dans l'Agriculture (FRIA-Belgium); the Agentschap voor Innovatie door Wetenschap en Technologie (IWT-Belgium); the Ministry of Education, Youth and Sports (MEYS) of the Czech Republic; the Council of Science and Industrial Research, India; the HOMING PLUS programme of the Foundation for Polish Science, cofinanced from European Union, Regional Development Fund; the OPUS programme of the National Science Center (Poland); the Compagnia di San Paolo (Torino); MIUR project 20108T4XTM (Italy); the Thalis and Aristeia programmes cofinanced by EU-ESF and the Greek NSRF; the National Priorities Research Program by Qatar National Research Fund; the Rachadapisek Sompot Fund for Postdoctoral Fellowship, Chulalongkorn University (Thailand); the Chulalongkorn Academic into Its 2nd Century Project Advancement Project (Thailand); and the Welch Foundation, contract C-1845.
\end{acknowledgments}

\bibliography{auto_generated}

\cleardoublepage \appendix\section{The CMS Collaboration \label{app:collab}}\begin{sloppypar}\hyphenpenalty=5000\widowpenalty=500\clubpenalty=5000\textbf{Yerevan Physics Institute,  Yerevan,  Armenia}\\*[0pt]
V.~Khachatryan, A.M.~Sirunyan, A.~Tumasyan
\vskip\cmsinstskip
\textbf{Institut f\"{u}r Hochenergiephysik der OeAW,  Wien,  Austria}\\*[0pt]
W.~Adam, E.~Asilar, T.~Bergauer, J.~Brandstetter, E.~Brondolin, M.~Dragicevic, J.~Er\"{o}, M.~Flechl, M.~Friedl, R.~Fr\"{u}hwirth\cmsAuthorMark{1}, V.M.~Ghete, C.~Hartl, N.~H\"{o}rmann, J.~Hrubec, M.~Jeitler\cmsAuthorMark{1}, A.~K\"{o}nig, M.~Krammer\cmsAuthorMark{1}, I.~Kr\"{a}tschmer, D.~Liko, T.~Matsushita, I.~Mikulec, D.~Rabady, N.~Rad, B.~Rahbaran, H.~Rohringer, J.~Schieck\cmsAuthorMark{1}, J.~Strauss, W.~Treberer-Treberspurg, W.~Waltenberger, C.-E.~Wulz\cmsAuthorMark{1}
% ...
\vskip\cmsinstskip
% ...
1:~~Also at Vienna University of Technology, Vienna, Austria\\
% ...

\end{sloppypar}
\end{document}